\documentclass[aps,prd,twocolumn,groupedaddress]{revtex4}
\usepackage{amsfonts}
\usepackage{eucal}
\usepackage{bm}
\usepackage{amssymb}
\usepackage{amsmath}
\usepackage[T1]{fontenc}
\newcounter{efn}
\newcommand{\scrif}{{\mathcal I}^{+}}

\newcommand{\T}{{\mathbb T}}

\newcommand{\N}{{\mathcal N}}
\newcommand{\Lie}{{\mathcal L}}

\newcommand{\scrC}{{\mathcal C}}

\newcommand\homi{{\hat\omicron}}
\newcommand\hio{{\hat\iota}}
\newcommand\cz{{\mathfrak z}}
\newcommand\cu{{\mathfrak u}}

\def\const{{\rm const}}

\def\geod{{\bm{\gamma}}}
\newcommand{\Spin}{{\mathbb S}}
\def\omicron{{o}}

\newcommand\ess{{\mathcal S}}
\begin{document}

\title{Estimating Energy--Momentum and Angular Momentum Near Null
Infinity}

\author{Adam D. Helfer}

\affiliation{Department of Mathematics, University of Missouri,
Columbia, MO 65211, U.S.A.}
\email{helfera@missouri.edu}

\date{\today}

\begin{abstract}
The energy--momentum and angular momentum contained in a spacelike
two-surface of spherical topology are estimated by joining the
two-surface to null infinity via an approximate no-incoming-radiation
condition.  The result is a set of gauge-invariant formulas for 
energy--momentum
and angular momentum which 
should be applicable to much numerical work; it also
gives estimates of the finite-size effects.
\end{abstract}

%numerical relativity

\maketitle

\section{Introduction}

The exciting successes in recent years of numerical treatments of
general-relativistic systems have given point to longstanding
theoretical challenges involved in explicating Einstein's theory.
Numerical relativists are now able to calculate with good
accuracy solutions to the field equations.  But how is one to extract
from all these data the key physically significant quantities?

Energy--momentum and angular momentum would be among the most
important, and the strongest hope is that good definitions of these
will exist at the {\em quasilocal} level, that is, on generic acausal
two-surfaces (perhaps restricted to have spherical topology).  If such
definitions do exist, and if, moreover, they can be formulated in such
a way that it is possible to meaningfully compare the quasilocal
quantities associated with different surfaces, they would be a
powerful tool for understanding the exchange of energy--momentum and
angular momentum between strongly generally relativistic astrophysical
systems.  At present, however, the problems involved in developing a
quasilocal kinematics seem so profound that one is driven to look for
more modest approaches which will still provide useful results in
broad categories of cases of current interest.

The goal of this paper is to get working approximations for the total
energy--momentum and angular momentum of isolated systems which are
suitable for contemporary numerical use.  Most numerical codes only
evolve the system throughout a finite volume of space--time, so one
has available data on large but finite two-surface $\ess$.  I shall
show that there is a reasonable ``poor man's'' no-incoming-radiation
condition which can be used to extrapolate the data on $\ess$ to
future null infinity $\scrif$, where the Bondi--Sachs definition of
energy--momentum and twistorial treatment of angular momentum apply.
It turns out --- nontrivially --- that this approach allows for a
comparison of the energy--momentum and angular momentum as $\ess$ is
moved forward in time; one thus has measures of the energy--momentum
and angular momentum emitted in gravitational radiation.

While a certain amount of work is required to derive these measures,
the result is a compact set of gauge-independent formulas which should
be usable by numerical relativists.

\subsection{A No-Incoming-Radiation Condition}

The state of a general-relativistic system is specified not only by
its material degrees of freedom but by its gravitational data.  In
principle, for most modeling one would like to fix those data to
correspond to no incoming radiation.  

To implement the no-incoming-radiation condition exactly would be very
difficult.  (One would have to solve a hard inverse problem, finding
what constraints on Cauchy data led to the required behavior of the
solution in the distant past.  Numerical workers typically specify the
data as well as they can before there are strong interactions, and
then discard any early transients as potentially due to spurious
incoming radiation.)  However, this is a practical difficulty and does
not affect the validity of the condition as the correct restriction on
the data.

We may make use of this observation at the two-surface $\ess$ (assumed
to be large and approximately spherical), as follows.  Consider the
null hypersurface $\N _{\rm phys}$ orthogonal outwards from $\ess$.
Radiation incoming to the future of $\ess$ would leave its profile on
this surface, the transverse component being measured by the Weyl
component $\Psi _0$ in the standard Newman--Penrose formalism.  
(One has $\Psi _0=\Psi _{ABCD}\omicron ^A\omicron ^B\omicron ^C
\omicron ^D$, where $\omicron ^A$  is a tangent spinor to $\N _{\rm 
phys}$, that is, one has $l^{AA'}=\omicron ^A\omicron ^{A'}$ for $l^a$ 
the null tangent to $\N _{\rm phys}$.)
We may
thus take as an approximate, ``poor man's,'' no-incoming-radiation
condition that
\begin{equation}\label{pmnirc}
\Psi _0 =0\mbox{ on the null hypersurface outgoing from }\ess
\end{equation}
(and there is no matter crossing this hypersurface).  More precisely,
we consider embedding $\ess$ in a space--time with the same first and
second fundamental forms, but 
we discard $\N _{\rm phys}$ and replace it with a null hypersurface 
$\N$ (orthogonally outwards from $\ess$)
with $\Psi _0$ identically zero on $\N$.  
(Note that condition~(\ref{pmnirc}) is imposed only on the single 
hypersurface $\N$ determined by $\ess$.  We shall take up below the 
question of what to do when $\ess$ is evolved.)

The spin-coefficient equations that effect transport outwards from
$\ess$ along $\N$ can be integrated, so one can work out the
asymptotic behavior of the fields as one approaches null infinity on
$\ess$.  One can then use the Bondi--Sachs energy--momentum and the
twistorial angular momentum. 

A number of remarks should be made about this.

First, according to the Sachs peeling property, it is the component
$\Psi _0$ has the most rapid fall-off along outgoing null geodesics
($\Psi _0\sim r^{-5}$ where $r$ is an affine parameter), and so one
has good reason to think that if $\ess$ can be regarded as in the
asymptotic regime, then $\Psi _0=0$ should be a good approximation.

Second, another way of viewing the nature of the approximation is
that, because $\N _{\rm phys}$ is at a finite location, the incoming
waves need not be exactly transverse.  That is, the extent to which
$\Psi _0=0$ fails to implement the exact no-incoming-radiation
condition is the extent to which incoming waves would not be
transverse at $\N _{\rm phys}$.  (This means in particular that the
approximation might be a poor one if $\ess$ were substantially
wrinkled.  However, in numerical work one typically has surfaces
$\ess$ which are very nearly spherical.  If one did have to deal with
substantially wrinkly surfaces, one could generalize the present
method by considering an outgoing null congruence other than the one
orthogonal to $\ess$, adapted to the ambient geometry.)

Finally, one might consider using the asymptotics of the field along
$\N$ for waveform extraction; cf. e.g.~\cite{BoyMrou}.  (Since these
asymptotics only give the field at an instant of retarded time
determined by $\ess$, to really extract a wave-form one would need to
apply the procedures here for an evolving family of surfaces $\ess 
(\eta
)$ with associated null hypersurfaces $\N (\eta)$, for $\eta$ in some
interval $J$.) Such a procedure would not be exact, of course; its
validity would be limited by the applicability of~(\ref{pmnirc}).  
While
this is certainly natural, it raises issues beyond those treated in 
this
paper.  This is because the points which are presently problematic in
the case of wave-form extraction and in the case of computation of
energy--momentum and angular momentum are different.  For
energy--momentum and angular momentum, resolving gauge ambiguities is
essential, and condition~(\ref{pmnirc}) allows us to do this (as a
mathematical procedure, irrespective of the degree to which the
condition accurately models the physical space--time); for wave-forms,
the gauge choices for the observer are trivial.  Thus the definitions
here allow one both to compute the energy--momentum and angular
momentum, and to estimate the finite-size effects involved, from data 
on
surfaces $\ess (\eta )$, but the problem of wave-form extraction
necessarily involves additionally the question of how accurately
condition~(\ref{pmnirc}) models the geometry of the physical
space--time.  One should also bear in mind that, for present work,
computations of (say) the energy--momentum or angular momentum emitted
in gravitational waves which were accurate to a few per cent would
typically be quite adequate; whereas extraction of physical 
information
from wave-forms may require rather more accurate modeling.  Because of
this sensitivity, if the ideas used for asymptotics here are to be
helpful to the wave-form extraction problem, they will probably be so
when combined with other physical insights.  One should probably use
data, not just on the surfaces $\ess (\eta )$, but from the portion of
the physical space--time interior to them, to extrapolate the
wave-forms.  

\subsection{The Role of Idealization}

The procedure used here turns on embedding the two-surface $\ess$ of
interest in a mathematically constructed space--time and then
evaluating that space--time's kinematics.  One might think at first
that this is less desirable than actually working out the asymptotics
along $\N _{\rm phys}$ to infer the Bondi--Sachs energy--momentum and
twistorial angular momentum in the physical space--time.  
For most
purposes, this is not the case, however.

The point is that the Bondi--Sachs treatment applies to idealized
isolated systems; for practical purposes one must choose which portion
of a real (or numerical) system is to be regarded as the isolated
component, and there may be several such choices.  Consider, for
instance, a system which at several widely separated intervals emits
bursts of gravitational radiation.  Each of these bursts contains, not
just the outgoing transverse wave front, but smaller trailing,
non-transverse, pieces; also each of these bursts will generate a
certain amount of back-scatter via nonlinearities.  If one wanted to
truly be in the mathematically exact Bondi--Sachs asymptotic regime
for this system, one would have to go outward so far along $\N _{\rm
phys}$ that one had passed any slight trailing fields and backscatters
due to very early emissions --- in principle, one would have to know
the history of the system in the arbitrarily distant past to do this.
One rarely wants to do this; one would rather think of the system as
to good approximation isolated in the interval around one burst ---
and if necessary then worry about the fact that the isolation is not
perfect.

Thus in most situations the task is not to construct the null infinity
and kinematics of the entire space--time, but to determine how to
measure the kinematics of a large but finite accessible region.   On
the other hand, the main reason for considering null infinity
(together with the Bondi--Sachs energy--momentum and the twistorial
angular momentum) --- that it provided an invariant treatment of
quantities of interest --- remains valid.  So our approach is based on
constructing a strictly well-defined null infinity from the data on
the finite surface $\ess$, in order to have an invariant
energy--momentum and angular momentum.
These are the energy--momentum and angular momentum
which would be ascribed to $\ess$, were it in a space--time 
satisfying~(\ref{pmnirc}).

The question of whether one really is in a regime which satisfies the
Bondi--Sachs asymptotics to a given approximation --- which is the
question of whether the energy--momentum and angular momentum
constructed here are stable when $\ess$ is moved outwards along $\N
_{\rm phys}$ --- is important but lies beyond the scope of this paper,
for one cannot investigate this stability based on data at a single
surface $\ess$ (nor from a finite family of $\ess (\eta )$'s).

\subsection{Contemporary Numerical Work}

Numerical relativists who work with codes aimed at treating generic
space--times have recognized the importance of invariant and
theoretically justified measures of energy--momentum and angular
momentum.  Indeed, this concern plays a part in the choice of some
groups to use characteristic (or mixed Cauchy--characteristic) codes.
Such codes, if cast in a form admitting a clean extension to null
infinity, ought to allow the extraction of the Bondi--Sachs
energy--momentum and twistorial angular momentum.  

Characteristic  (or mixed) code computations of the Bondi energy
(often referred to in the literature as the Bondi mass, for historical
reasons) have been done in a number of cases, although typically they
are hampered by the fact that the codes themselves are not usually
cast in Bondi coordinates, so that non-trivial gauge transformations
are required.  These also make the computation of the Bondi momentum
and the angular momentum difficult.  (See ref.~\cite{Winicour2009} for
a review.)  The results here may help streamline such computations,
for the formalism to be developed automatically produces the required
gauges.

A great body of numerical work, however, is based on ``$3+1$''
formalisms rather than characteristic or mixed ones.  For these
formalisms, the problem of gauge invariance for energy--momentum and
angular momentum has been more severe.

Most contemporary attempts to extract information about the total
energy--momentum in the $3+1$ formalisms can be usefully thought of as
based on the Bondi--Sachs energy--momentum loss formula \footnote{When
the objects of interest are black holes, there are also some
specialized techniques for estimating their masses.  However, these
rely strongly on stationarity.  The problem of identifying the
corresponding energy--momenta as elements of a suitable space in the
asymptotic regime is unaddressed, too, so one does not have a good way
of comparing the energy--momenta of several interacting holes, or even
one radiating one at different times}.  (They have often been
justified by other means; however, the Bondi--Sachs formula would be
the broadest and most theoretically secure starting-point.) This
formula identifies the rate of change of the Bondi--Sachs
energy--momentum, with respect to Bondi retarded time, as an integral
of the squared modulus of the ``news function'' with respect to the
measure induced on an asymptotically large sphere by the Bondi
coordinates.  Thus to use this formula to recover the radiated
energy--momentum one must know:  (a) that one is in the asymptotic
regime; (b) the Bondi coordinate system, both to identify the measure
on the sphere correctly and to do the integral over retarded time; (c)
the news function.  (The news function can be given as the integral of
a curvature component --- the component depending on the Bondi
coordinates --- with respect to Bondi retarded time.)

Numerical work in the $3+1$ formalism has not yet implemented any
systematic transition to Bondi coordinates.  Thus what is actually
done is to use the numerical angle and time coordinates to compute the
curvature component and integrals required for the 
energy--momentum-loss formula and its integral  (e.g.
\cite{LZ2008,Pretorius2009}).  Consistency checks are then done by
studying the stability of the result as the extraction radius is
increased.  However, the lack of gauge control makes it impossible to
know what these numbers really signify.  The stability of the
actual, gauge-invariant,
energy--momentum could be either better or worse than the cited
numbers, depending on whether the gauge freedoms exacerbate or mask
extraction problems.

The present approach overcomes the concerns about gauge by giving
formulas for the Bondi--Sachs energy--momentum and its evolution in
terms of gauge-invariant quantities on the extraction surface $\ess$.
As far as the question of the stability of the results with increasing
extraction radius goes, this is an issue which one can only
investigate directly, by considering larger and larger surfaces.
However, the present work does give one the confidence that in such an
investigation other potential error sources have been controlled.

The situation for angular momentum has been more difficult than for
energy--momentum.  In the first place, there has been for some time no
really theoretically satisfactory formula; and in the second, 
the angular momentum depends on curvature terms deeper in
the asymptotic expansion, which are still more sensitive to the
correct choice of Bondi frame.  The approach given here
overcomes these difficulties.  We use the recent twistorial definition
of angular momentum, which appears to be theoretically satisfactory.
We identify the Bondi frame exactly on $\N$, and the finite-size
contributions may be read off directly from the formulas here.

\subsection{Recent Theoretical Work}

The problem of estimating kinematic quantities in terms of numerical
data has been taken up by two sets of authors recently.

Gallo, Lehner and Moreschi~\cite{GLM2008} (see also \cite{GLM2007})
raised many of the concerns motivating the present work.  They
emphasized the importance of extracting invariant information, and
also of considering finite-size effects.  They gave an approach to
estimating the Bondi momentum which (while presented somewhat
differently) can be viewed as assuming that the two-surface $\ess$ is
only infinitesimally separated from $\scrif$, and computing the
Bondi--Sachs energy--momentum at the corresponding cut.  (While it may
seem odd to speak of a two-surface infinitesimally separated from
infinity, it is a well-defined concept from the point of view of the
conformally completed manifold, and amounts to assuming that one only
needs leading terms in the appropriate asymptotic expansions.)  Their
results therefore correspond to a limiting case of some of those here. 

Deadman and Stewart~\cite{DS2009} recently discussed the estimation of
the Bondi--Sachs energy from numerical data.  Their approach is rather
different; it is based on constructing a transformation from the
coordinates of the numerical evolution to Bondi-like coordinates.  As
these formulas are deduced by considering finitely many terms in the
asymptotic expansions and extrapolating, this work could also be 
viewed
as based on the notion that $\ess$ was infinitesimally separated from
$\scrif$.

\subsection{Some Technical Points}

A significant feature of the present approach is that, with it, all
the quantities of interest can be expressed in terms of standard
Geroch--Held--Penrose (the boost-weight-covariant version of the
Newman--Penrose calculus) quantities at $\ess$, and the
energy--momentum and angular momentum are given as natural integrals
of these at $\ess$.  The formulas derived here include within
themselves all necessary changes to refer to Bondi frames; no separate
computation of Bondi coordinates is necessary.

The question of how accurately the Geroch--Held--Penrose quantities
can be computed at $\ess$ of course depends on the particular code.
Presumably the most difficult one to measure accurately is $\Psi _1$,
which is central in computing the angular momentum (and also, because
of finite-size effects, contributes to the energy--momentum).  It
should be emphasized that the question here is only that of the
computation of $\Psi _1$ at $\ess$, not of its inferred asymptotic
value (an issue raised by Deadman and Stewart~\cite{DS2009}).

A second point is that we need not take up the delicate questions of
just what degree of smoothness or peeling is encoded in the numerical
solution.  This is because we have separated the question of computing
the energy--momentum and angular momentum from the question of finding
their limiting values at $\scrif$:  our results are given entirely in
terms of data at $\ess$. 

The approach here also allows one to quantify finite-size-effects, and
so provides a useful consistency-check on the degree to which $\ess$
is ``effectively at infinity'' (using only data at $\ess$).  The
integrals for the energy--momentum and angular momentum at null
infinity are given in terms of the asymptotic values of the curvature
quantities $\Psi _1$, $\Psi _2$, $\Psi _3$; here, those asymptotic
values appear as the values at $\ess$ (suitably scaled) plus
correction terms.  Those correction terms, then, are a measure of how
removed $\ess$ is from null infinity.  We also, as importantly, are
able to quantify how strongly the inferred structure of null infinity
is subject to finite-size effects as evolution proceeds.

In numerical work, the surfaces $\ess$ are typically large coordinate
spheres, and their first and second fundamental forms appear as slight
perturbations of the values they would have for large spheres in
Minkowski space.  In particular, the convergence $\rho _\ess$ of the
outgoing congruence is a slight perturbation of $-R^{-1}$ (by
$O(R^{-2})$ or less), and the shear $\sigma _\ess$ is expected to be
$O(R^{-2})$ if there is no incoming radiation; here $R$ is the radius
of the sphere.  Thus to good approximation 
\begin{equation}
|\sigma _\ess |\ll {\left|\rho _\ess \right|}\, .
\end{equation}
(If there were incoming radiation, one would expect $\sigma _\ess$ to
go like a dimensionless number --- the news function of that radiation
--- over $R$.  Thus a small amount of incoming radiation would not
upset this inequality.) This is helpful, for the asymptotic forms of
$\Psi _2$, $\Psi _3$ on $\N$ are simplified in this case, and we make
this approximation in computing them.  If more accuracy is needed for
particular work, probably the most efficient approach would be to
compute $\Psi _2$, $\Psi _3$ perturbatively in $\sigma _\ess /\rho
_\ess$ to the required order.  (In such computations, note that while
$|\sigma _\ess |\ll \left|\rho _\ess \right|$, the angular derivatives 
of $\rho
_\ess$ and $\sigma _\ess$ may very well be of the same size.  Thus one
must be careful not to discard at one stage terms whose angular
derivatives may be essential later.) Curiously, while $\Psi _2$, $\Psi
_3$ are very complicated, all other elements of the calculation are
manageable; in particular, even the exact forms of the asymptotic
spinors and twistors are simple.

\subsection{Evolution}

I have so far emphasized that, by casting the problem of measuring
energy--momentum and angular momentum at a finite two-surface in a
certain form, one can defer the difficult questions surrounding
quasilocal kinematics.  However, these questions must be faced to some
degree when we consider the evolution of the system, and the question 
of
how to compare the energy--momenta and angular momenta at two 
different
two-surfaces.  This is because the auxiliary space--times constructed
via the poor man's no-incoming-radiation condition from the two 
surfaces
are not the same, and so it is not obvious how to identify the spaces 
on
which their energy--momenta and angular momenta take values.  Indeed, 
we
must anticipate on physical grounds that unless the extraction 
surfaces
are large enough there will be no way of identifying their auxiliary
null infinities which preserves all of the usual structures.  

This is an instance of a more general problem for quasilocal
kinematics:  how is one to compare the kinematic quantities associated 
with different two-surfaces?  Quasilocal kinematic proposals are not 
well-enough developed at present to take up this problem, but the 
degree to which quasilocal kinematics will be useful depends very 
largely on the degree to which it can be solved.

It turns out that, for us, there is a natural approach to this problem
which fits well with structures previously developed for treating
angular momentum at null infinity.   We shall see that there is a
natural way to compare the null infinities from two surfaces $\ess$,
$\ess '$ infinitesimally separated in time; this procedure can then be
integrated.  Potential finite-size effects show up in that the
identifications of the null infinities are not via
Bondi--Metzner--Sachs transformations, but via more complicated
motions, unless the extraction surfaces are large enough.   Thus one
can say that while a single null infinity is being used, to the extent
that finite-size effects are important the null infinity has a weaker
structure than is conventional.

More precisely, we suppose we have a one-parameter family of 
two-surfaces $\ess (\eta )$ (for $\eta$ in some interval $J$)
foliating a timelike three-surface $T$, with $\eta$ increasing towards 
the future.
Each two-surface $\ess (\eta )$ is embedded (with the same first and 
second fundamental forms)
in an auxiliary space--time 
$M(\eta )$ defined by taking $\Psi _0=0$ along a null surface 
$\N (\eta )$ orthogonally outwards from $\ess (\eta )$.  Then the 
constructions already described give a null infinity $\scrif (\eta )$ 
for each $M(\eta )$, with the null geodesics orthogonally outwards 
from $\ess (\eta )$ defining a preferred cut $\scrC (\eta )\subset
\scrif (\eta )$.  Thus really we have a bundle of space--times 
$\{ M(\eta )\mid\eta \in J\}$.\footnote{One technicality 
may be worth mentioning:  As usual, when one writes of ``preserving 
the second fundamental form'' under an embedding (here 
$\ess (\eta )\to M(\eta )$), one really means that the embedding is 
accompanied by an isomorphism of the ambient tangent spaces (here
$\phi (\eta ):
T(M(\eta))\Bigr| _{\ess (\eta )}\to T(M_{\rm phys})\Bigr| _{\ess 
(\eta )}$, an isomorphism of spatially and temporally oriented 
Lorentzian vector bundles).
Thus strictly speaking we 
consider, for each $\eta\in J$, not just the space--time $M(\eta )$ 
but a pair $(M(\eta ),\phi (\eta ))$, and we should work with the 
bundle 
$\{ (M(\eta ),\phi (\eta ))\mid \eta\in J\}$ of such pairs.  However, 
we shall not need such an explicit 
formalism.}\setcounter{efn}{\thefootnote}
(We do {\em not} have a single 
space--time for which the condition $\Psi _0=0$ holds on a local 
foliation of null surfaces; that would generally be far too 
restrictive a condition to impose.)

With these structures, it turns out that there is a natural way to 
identify the null infinities $\scrif (\eta )$ for the different $\eta
$'s.  
The key step is to make the identifications at an infinitesimal level; 
one can then integrate. 
The main issue then comes down to 
understanding how one should define the cut of 
$\scrif (\eta )$ corresponding to the null geodesics orthogonally 
outwards from the two-surface
$\ess (\eta +d\eta )$, that is, {\em for $\eta +d\eta$ infinitesimally 
differing from} $\eta$.  Once this is done, one can fix both the 
identification of the generators (because one gets a point-to-point 
mapping of the cuts) and the supertranslation freedom.

The actual identification we need arises from natural isomorphisms.  
Let us begin with a vector field $w^a$ in $M_{\rm phys}$ connecting $
\ess (\eta )$ to $\ess (\eta +d\eta )$.  (That is, the field $w^a$ is 
tangent to $T$ and $w^a\nabla _a\eta =1$.)  At each point of $\ess 
(\eta )$, we may consider the Jacobi field along the outgoing null 
geodesic in $M_{\rm phys}$ whose initial value is $w^a$ at that point 
(and whose initial velocity is chosen to make the field represent a 
null geodesic outwards orthogonal to $\ess (\eta +d\eta )$).  We thus 
get a family of Jacobi fields, over $\ess (\eta )$, which represent 
null geodesics orthogonally outwards from 
$\ess (\eta +d\eta )$.  All of this so far is in $M_{\rm phys}$.

Now, any Jacobi field is specified by its initial data.
By the construction of $M(\eta )$, there is a natural isomorphism of 
the tangent bundles $T(M_{\rm phys})\Bigr| _{\ess (\eta )}\cong 
T(M(\eta ))\Bigr| _{\ess (\eta )}$.  (See footnote~[\theefn].
The symbol $T$ which occurs as part of the notation $T(X)$
the tangent bundle of $X$ should not be confused with the isolated $T$ 
representing the timelike three-surface foliated by the $\ess (\eta )
$'s.)
Thus we may naturally identify the Jacobi fields we found above with 
fields {\em in the auxiliary space--time} $M(\eta )$.  The limiting 
values of these fields at $\scrif (\eta )$ we take to define the 
displacement of the cut corresponding to $\ess (\eta +d\eta )$ from $
\scrC (\eta )$.  (This will be well-defined independent of questions 
about the asymptotic behavior of the original Jacobi fields in the 
physical space--time.)  It is this definition we required for the 
identification of $\scrif (\eta )$ with $\scrif (\eta +d\eta )$, and 
its subsequent integration to give identifications of 
the null infinities $\scrif (\eta ' )$ for different values of $\eta 
'$.
 
As noted above, these identifications will not be perfect 
Bondi--Metzner--Sachs motions, because of finite-size effects.  The 
formulas we derive for the evolution of energy--momentum and angular 
momentum apply even in this case.
However, for purposes of extracting the {\em
total} energy--momentum and angular momentum, substantial finite-size
effects (that is, non-Bondi--Metzner--Sachs identifications) should be
regarded as signaling that the extraction surfaces have not been taken
to be distant enough.  The finite-size results are more of interest at
present in that they may provide clues about how to develop quasilocal
kinematics generally.

Having discovered that there is a well-defined null infinity
(if with somewhat weaker than usual properties) for the family of 
extraction surfaces $\{ \ess (\eta )\mid \eta \in J\}$, it is natural 
to ask if one cannot construct a single asymptotic regime for the 
physical space--time $M_{\rm phys}$ to which this null infinity is 
attached?  In some sense, this is provided by the bundle $\{ \N 
(\eta )\mid \eta\in J\}$, which can be attached to $M_{\rm phys}$ 
along 
$T$.  However, this bundle is not usually a space--time (it does not 
admit a metric structure compatible with the geometry of the $\N 
(\eta )$'s, since the condition $\Psi _0=0$ is generally
too strong to impose on a foliating family of hypersurfaces).  
While such constructions might be of some interest in the general 
problem of defining asymptotic regimes, in this paper there will be no 
reason to make use of them; considering the individual auxiliary 
space--times $M(\eta )$ and the identifications of their null 
infinities will be what is relevant.

\subsection{Outline and Conventions}

The next section of this paper outlines the integration of the
Newman--Penrose equations under the ``poor man's''
no-incoming-radiation condition.  In Section~III, the asymptotic
reference frames of Bondi and Sachs are introduced.  Section~IV gives
the computation of asymptotically constant spinors and the kinematic
twistor.  Section~V gives the Bondi-Sachs energy--momentum, and
Section~VI the twistorial angular momentum.  Section~VII derives the
formulas for comparing the energy--momentum and angular momentum (as
well as the deformation of the numerical coordinates relative to the
Bondi coordinates) as the two-surface is moved forward in time.
Section~\ref{UG} is a Users' Guide to the results; it summarizes what
procedures and equations would be needed in numerical work.  

A reader wishing simply to use the results here can use 
Section~\ref{UG} as an index to the paper.  
(Since the twistorial
angular momentum is new, the reader wishing to use this
will probably want to read
the introduction to Section~VI and also VI~A.)

All material necessary for understanding this paper and not otherwise
cited can be found in ref.~\cite{PR1986}, whose notation and 
conventions
are used.  Ref.~\cite{ADH2007} gives the twistorial treatment of
angular momentum at $\scrif$; ref.~\cite{ADH2009} is an account of it
for non-specialists.

The standard literature uses the symbol $\lambda$ for three different
things:  a spin-coefficient, a rescaling factor for the spinor dyad,
and the angular potential for the shear.  We shall use $\lambda _{\rm 
B}$ for
the angular potential for the shear and $\lambdabar$ for the rescaling
$\omicron ^A\mapsto\lambdabar\omicron ^A$.  (The spin-coefficient will
be $-\sigma '$; it will play little explicit role in this paper.)

The metric signature is $+{}-{}-{}-$.  The symbol $\sim$ will be used
to denote asymptotic equality as one moves outwards along null
geodesics.  The symbol $\approx$ stands for equality modulo $o(|\sigma
_\ess |/|\rho _\ess |)$.  All logarithms are natural.

\section{Asymptotics from Local Data}

The aim here is to compute those asymptotic quantities we will need
--- enough to find the energy--momentum and angular momentum at
$\scrif$ along $\N$ --- in terms of local data on $\ess$.  This can be
reduced to a series of integration problems.  Some of
the details of the (lengthy, but straightforward) integrations are
omitted.

What we require are the asymptotic forms of the curvature quantities
$\Psi _1$, $\Psi _2$, $\Psi _3$, as well as the operators $\eth$,
$\eth '$ and the shear $\sigma$.  (We also must verify that certain
other spin coefficients have an appropriate asymptotic decay, even if
we do not use their values; however, these decays will be obvious from
the general forms of the integrals determining them, except in one
case, which we shall treat explicitly.) And we must identify the
correct Bondi frame, that is, we must be sure that when we take the
asymptotic limits, by looking at two-surfaces receding to infinity
in null directions orthogonally outwards from
$\ess$, their null inward normals are compatible with those used
in the standard analysis.  

In this section, the subscript $\ess$ is used to indicate the value of
a quantity at $\ess$ rather than at an arbitrary point of $\N$.
However, having found the expressions for all quantities of interest
in terms of data at $\ess$, in later sections almost all computations
will be expressed in terms of these data, and the subscript will
be omitted.

\subsection{The Integration Scheme}\label{intsch}

Let $\ess$ be a spacelike two-surface of spherical topology in a
vacuum region of space--time, and let $\omicron ^A$, $\iota ^A$ and
$l^{AA'}=\omicron ^A\omicron ^{A'}$, $m^{AA'}=\omicron ^A\iota ^{A'}$,
${\overline m}^{AA'}=\iota ^A\omicron ^{A'}$, $n^{AA'}=\iota ^A\iota
^{A'}$ be a spinor dyad and vector tetrad associated with it, so $l^a$
is the outgoing null congruence and $n^a$ is the ingoing null
congruence.  Assume next that this surface were embedded in a
space--time with the same first and second fundamental forms, and with
the same values of $\Psi _n$ for $1\leq n\leq 4$ at $\ess$.
(Actually, the value of $\Psi _4$ will not enter.) We further assume
that $\omicron ^A$, $\iota ^A$ are propagated parallel along the
outgoing null congruence from $\ess$, and that $\Psi _0$ vanishes
along this congruence.  This leaves a freedom $\omicron
^A\mapsto\lambdabar\omicron ^A$, $\iota ^A\mapsto\lambdabar ^{-1}\iota
^A$ where $\lambdabar$ is a function of the generator only.

We have then that $\rho =\overline\rho$ on the null congruence, and
also that $\rho '_\ess ={\overline{\rho '}}_\ess$, where the subscript
indicates restriction to $\ess$.  The conditions for $\omicron ^A$,
$\iota ^A$ to be propagated parallel along $l^a$ mean the
spin-coefficients $\epsilon$, $\kappa$, $\gamma '$ and $\tau '$ all
vanish.  The choices so far apply to the spin-frame on $\N$, but it
will also be convenient to restrict its behavior to first order off
$\N$.  We require that $\nabla _{[a}l_{b]}=0$, which implies $\tau
=\overline\alpha +\beta$.

The restrictions $\epsilon=0$, $\gamma '=0$, and $\tau
=\overline\alpha +\beta$ break the strict boost invariance of the
Geroch--Held--Penrose calculus.  However, a modified invariance still
holds.  If we consider rescaling $\omicron ^A\mapsto
\lambdabar\omicron ^A$, $\iota ^A\mapsto\lambdabar ^{-1}\iota ^A$,
where $\lambdabar$ is a non-zero complex-valued function on $\N$ which
depends on the generator only, then the conditions $\epsilon =0$,
$\gamma ' =0$ are preserved.  If we consider that, accompanying any
such rescaling the spinor field $\omicron ^A$ is changed to first
order off $\N$ by $D'\omicron ^A\mapsto
{\overline\lambdabar}^{-1}(D'\omicron
^A-{\overline\lambdabar}^{-2}(\eth \lambdabar\overline \lambdabar)
\iota ^A )$, then we find $\tau\mapsto \lambdabar
{\overline\lambdabar}^{-1}(\tau +(\lambdabar\overline\lambdabar
)^{-1}\eth \lambdabar\overline\lambdabar )$, $\overline\alpha
+\beta\mapsto \lambdabar {\overline \lambdabar}^{-1}(\overline\alpha
+\beta )+{\overline\lambdabar}^{-2}\eth \lambdabar\overline\lambdabar$
and $\tau =\overline\alpha +\beta$ is preserved.   Since this modified
transformation law for the spinor dyad differs from the simple
rescaling of the GHP scheme only by altering the derivative
$D'\omicron ^A$ by a multiple of $\iota ^A$, the only spin-coefficient
inhomogeneously affected on $\N$ is $\tau$; also the operators $\eth$,
$\eth '$ on $\N$ retain their usual GHP transformation rules.  (The
behavior of $\tau$ adopted here is natural within the context of the
characteristic initial-value problem;
cf.~\cite{Sachs1962,Friedrich1981}.)

Let $s$ be an affine parameter along the outgoing null geodesics
normalized so that $l^a\nabla _as=1$ and vanishing at $\ess$.  We may
think of $s$ having Geroch--Held--Penrose type $\{ p,q\} =\{ -1,-1\}$
(given our restrictions $D\omicron ^A =0$, $D\iota ^A=0$).  (It is
conventional to use $r$ for an affine parameter if the boost freedom
in the spin-frame is fixed to give the standard Bondi--Sachs
asymptotics.  However, as we shall have not fixed the spin frame in
this way, we use $s$ to avoid potential confusions.)

All of the computations are built on the integration of the optical
equations, which can be written as
\begin{equation}
D\left[\begin{array}{cc} \rho &\sigma\\ 
   \overline\sigma &\rho\end{array}\right]
  =\left[\begin{array}{cc} \rho &\sigma\\ 
   \overline\sigma &\rho\end{array}\right]^2\, ,
\end{equation}
since $\Psi _0=0$ and $\Phi _{00}=0$.  The solution to this is 
\begin{equation}\label{optsol}
\left[\begin{array}{cc} \rho &\sigma\\ 
   \overline\sigma &\rho\end{array}\right]
 =\left( -s\left[\begin{array}{cc} 1 &0\\ 
     0 &1\end{array}\right] + \left[\begin{array}{cc} \rho &\sigma\\ 
     \overline\sigma &\rho\end{array}\right] _\ess^{-1}
\right) ^{-1}\, .
\end{equation}
Notice that these matrices, for different values of $s$, all commute,
since they lie in the commutative algebra generated by the identity
and the single matrix $\left[\begin{array}{cc} 0 &\sigma\\
\overline\sigma &0\end{array}\right] _\ess$.  It is this which allows
the integration of the system explicitly.

It is easy to see that $\rho$ and $\sigma$ will be non-singular for
$s\geq 0$ iff $\rho _\ess < -|\sigma _\ess|\leq 0$, which is the
condition that no conjugate points develop.  In practice one expects
$\rho _\ess$ to fall off as $1/r$ and $\sigma _\ess $ to fall off as
$1/r^2$ where peeling holds (where $r$ is the affine parameter in the
physical space--time, not the mathematical stand-in with $\Psi _0=0$),
so for large enough surfaces $\ess$ which are close enough to spheres
one should have $|\sigma _\ess|\ll \left|\rho _\ess \right|$.  This 
means that
for such surfaces it should be a good approximation to neglect the
effects of $\sigma$ relative to $\rho$ on propagation outwards.
In this approximation, we have
\begin{equation}\label{rhosigmaeqn}
\left[\begin{array}{cc} \rho &\sigma\\ \overline\sigma &\rho
\end{array}\right]
\approx
\left[\begin{array}{cc} 
  \rho_\ess (1-s\rho _\ess )^{-1} 
    &\sigma _\ess(1-s\rho _\ess)^{-2}\\ 
  {\overline\sigma}_\ess (1-s\rho _\ess)^{-2} &
  \rho _\ess (1-s\rho _\ess )^{-1} \end{array}\right]\, .
\end{equation}
Here and throughout, we shall use $\approx$ to indicate the
condition $|\sigma _\ess|\ll \left|\rho _\ess \right|$.  

The remaining equations are integrated successively, as follows.  (The
results are given below, Tables~\ref{firsttable}, \ref{secondtable}.)  
From the equation
\begin{equation}\label{Psi1eq}
 D\Psi _1=4\rho \Psi _1
\end{equation}
one finds $\Psi _1$.  With that in hand, one takes up that for the
spin-coefficients $\alpha$ and $\beta$, which we write as
\begin{equation}
D\left[\begin{array}{c}\alpha\\ \beta\end{array}\right]
  =\left[\begin{array}{cc} \rho &\overline\sigma \\ 
       \sigma &\rho\end{array}\right]
  \left[\begin{array}{c}\alpha\\ \beta\end{array}\right]
  +\left[\begin{array}{c}0\\ \Psi _1\end{array}\right]\, .
\end{equation}
This, and others to follow, can be integrated using the result
\begin{equation}\label{intfac}
\exp\int _0^s\left[\begin{array}{cc} \rho &\overline\sigma\\
  \sigma &\rho\end{array}\right] (s')\, ds'
  =\left[\begin{array}{cc} \rho &\overline\sigma\\
  \sigma &\rho\end{array}\right] 
  \left[\begin{array}{cc} \rho &\overline\sigma\\
  \sigma &\rho\end{array}\right] ^{-1}_\ess\, .
\end{equation}
(Recall that the matrices in the exponential all commute, so there is
no need to take a path-ordered exponential.)

One next takes up the transport of the operators $\delta$, $\delta '$
up the generators.  
Lie transport along $l^a$, the vector tangent to these, establishes a
canonical
diffeomorphism of the outgoing null surface $\N$ 
with $\ess\times \{ s\mid s\geq 0\}$.  Using this diffeomorphism, we may 
extend $\delta _\ess =m^a_\ess\nabla _a$ 
from its definition on $\ess$ to $\N$; this is equivalent to extending 
it by requiring it to be Lie transported along $l^a$.
If we put $m^a_\ess =Am^a +B{\overline m}^a+Cl^a$, then using
the standard spin-coefficient commutators we find the conditions for
being Lie-transported are 
\begin{equation}
D\left[\begin{array}{c} A\\ B\\ C\end{array}\right]
  =-\left[\begin{array}{ccc}\rho &\overline\sigma &0\\
  \sigma &\rho &0\\ -(\beta +\overline\alpha ) &-(\overline\beta
+\alpha ) &0\end{array}\right]
  \left[\begin{array}{c} A\\ B\\ C\end{array}\right]\, ,
\end{equation}
which can be integrated using~(\ref{intfac}) and the initial
conditions $A=1$, $B=0$, $C=0$.

With the results of these integrations, one can find the remaining
quantities of interest.  For $\Psi _2$, we integrate
\begin{equation}
D\Psi _2=(\delta ' - 2\alpha )\Psi _1+3\rho\Psi _2\, .
\end{equation}
With this, we can integrate the transport equation for the optical
scalars of the ingoing congruence:
\begin{equation}
D\left[\begin{array}{c}\rho '\\ \sigma '\end{array}\right] =
\left[\begin{array}{cc} \rho &\sigma\\ \overline\sigma
&\rho\end{array}\right]
  \left[\begin{array}{c}\rho '\\ \sigma '\end{array}\right]
  +\left[\begin{array}{c} -\Psi _2\\ 0\end{array}\right]\, .
\end{equation}
In fact, of these, we only need $\sigma '$ as a datum for the next
equation, and we need to note that $\rho '$ and $\sigma '$ will have,
by virtue of~(\ref{intfac}), the asymptotic behavior $O(s^{-1})$.
Then one can integrate
\begin{equation}
  D\Psi _3=\delta '\Psi _2+2\rho \Psi _3 +2\sigma '\Psi _1\, .
\end{equation}

\begin{table*}[!]
\caption{\label{firsttable}Asymptotic forms of the relevant
spin-coefficients, coefficients of operators, and curvature,
in the parallel-transported frame, in the case
$|\sigma _\ess |\ll |\rho _\ess |$.  
The left-hand column gives the asymptotic form of each quantity
$X$ for large affine
parameter $s$ in terms of a leading coefficient $X^0$; the coefficient
is given in the right-hand column.
In the right-hand column, all
quantities are to be evaluated at $\ess$ (and so would, in the 
notation of this Section, ordinarily 
carry the subscript $\ess$, but this is omitted here).}
\begin{ruledtabular}
\begin{tabular}{cc}
Quantity $X$ &Leading coefficient $X^0$, assuming
$|\sigma _\ess|\ll |\rho _\ess|$\\
$\alpha\sim\alpha ^0s^{-1}$&$(-\rho )^{-1}\alpha$\\
$\beta\sim\beta ^0 s^{-1}$&$(-\rho )^{-1}
 \left\{\beta -(2\rho )^{-1}\Psi _1\right\}$\\
$A\sim A^0s$&$-\rho $\\
$B\sim B^0 s$&$0$\\
$C\sim C^0s$&$\overline\alpha +\beta $\\
$\rho\sim \rho ^0 s^{-1}$&$-1$\\
$\sigma\sim\sigma ^0 s^{-2}$&$(-\rho )^{-2}\sigma$\\
$\Psi _1\sim \Psi _1^0 s^{-4}$& $(-\rho )^{-4}\Psi _1$\\
$\Psi _2\sim\Psi _2^0 s^{-3}$& $(-\rho )^{-3}\left\{
  \Psi _{2} -\rho ^{-1}\eth '\Psi _{1} +2\rho ^{-2}\Psi _{1}\eth '\rho 
   +2(3\rho )^{-2}|\Psi _{1}|^2\right\}$\\
$\Psi _3\sim \Psi _3^0s^{-2}$&\parbox{50em}{\begin{eqnarray*}
(-\rho )^{-2}\left\{\Psi _{3}+\rho ^{-1}\sigma '\Psi _1\right.
 &+&(1/6)\eth ' [-4\rho ^{-3}\Psi _{1}\eth '\rho
 +3\rho ^{-2}\eth '\Psi _{1} -\rho ^{-3}|\Psi
_{1}|^2 -\rho ^{-1}\Psi _{2}]\\
&+&(30\rho ^{4} )^{-1}(\eth '\rho )[15\Psi _{1}\eth '\rho 
  -10\rho\eth '\Psi _{1} +4|\Psi
_{1}|^2 +15\rho ^2\Psi _{2}]\\
&+&\left. (30\rho ^4)^{-1}[15|\Psi _{1}|^2\eth '\rho +4\Psi
_{1} (\overline{\Psi _{1}})^2 -10\rho \Psi _{1}\eth
'\Psi _{1} +15\rho ^2\overline{\Psi _{1}} \Psi
_{2}]\right\}\end{eqnarray*}}
\end{tabular}
\end{ruledtabular}
\end{table*}

Recall that we are interested in the asymptotic forms of
those spin-coefficients and operators
necessary to define the asymptotic spinors and twistors, and also the
curvature quantities $\Psi _1$, $\Psi _2$, $\Psi _3$. 
Table~\ref{firsttable} gives these results.  The
asymptotic values of the quantities $\alpha$, $\beta$, $A$, $B$, $C$
enter into the determination of the asymptotic spinors and twistors,
and there is little point in discussing them before those results are
at hand.  The asymptotic values of $\Psi _1$, $\Psi _2$, $\Psi _3$ are
of some interest, however.  For $\Psi _2$ and $\Psi _3$, we see a
leading term which is the value on $\ess$ multiplied by the
power of $(1-s\rho )$ dictated by peeling, plus correction terms.
One can check that, if the usual peeling assumptions hold in the 
physical space--time, then the scaling of these correction terms with 
the
position of $\ess$ is subdominant to that of the leading term.  Thus 
the terms represent finite-size effects due
to the distance of $\ess$ from $\scrif$, and, despite their 
complexity,
should be small for large enough $\ess$.  The complicated nonlinear
forms of these corrections are also of some interest.

While we expect the results of Table~\ref{firsttable} to suffice for
most numerical work, it is of some conceptual interest to extend these
by dropping the assumption $|\sigma _\ess |\ll {|\rho _\ess|}$.  This 
is
done for those quantities needed to determine the asymptotic spinors
and twistors in Table~\ref{secondtable}.  Expressions for the
curvature quantities $\Psi _2$ and $\Psi _3$ 
in this case are prohibitively lengthy; if any
are needed, the best approach is to do the corresponding integrals as
power series in $\sigma _\ess /\rho _\ess$, keeping as many terms as
required.

\begin{table*}
\caption{\label{secondtable}Exact leading spin-coefficients and
coefficients of operators in the parallel-transported frame.  In the
right-hand column, all quantities are 
to be evaluated at the surface
$\ess$ (and so would, in the notation of this Section, ordinarily 
carry the subscript $\ess$, but this is omitted here).}
\begin{ruledtabular}
\begin{tabular}{cc}
Coefficient&Value in terms of data at $\ess$\\
\hskip-2em$\left[\begin{array}{c}\alpha ^0\\ \beta ^0\end{array}
\right]$
 &\hskip-4em\parbox{50em}{\[ -(\rho ^2-|\sigma |^2 )^{-1}
   \left[\begin{array}{cc}\rho&-\overline\sigma\\
-\sigma&\rho\end{array}\right]\left\{
\left[\begin{array}{c}\alpha\\ \beta\end{array}\right]
 +\left(\frac{\rho ^2-|\sigma |^2}{4|\sigma |^3}
\log\frac{\rho+|\sigma |}{\rho -|\sigma |} -\frac{\rho}{2|\sigma
|^2}\right)\left[\begin{array}{c}0\\ \Psi _1\end{array}\right] 
  -\left(\frac{\rho}{4|\sigma |^3}\log\frac{\rho +|\sigma
|}{\rho-|\sigma |} -\frac{1}{2|\sigma |^2}\right)
\left[\begin{array}{c}\overline\sigma\\ \rho\end{array}\right]\Psi
_1\right\}\]}\\
\hskip-2em $\left[\begin{array}{c}A^0\\ B^0\end{array}\right]$
 &$-\left[\begin{array}{c}\rho\\ \sigma\end{array}\right]$\\
\hskip -2em $C^0$
 &\parbox{50em}{\[ 
    \left\{ \overline\alpha +\beta -(\rho\Psi _1 +\sigma
\overline{\Psi _1})\left(\frac{\rho}{4|\sigma |^3}\log\frac{\rho
+|\sigma |}{\rho -|\sigma |} -\frac{1}{2|\sigma |^2}\right)
  +\left(\frac{\rho ^2-|\sigma |^2}{4|\sigma
|^3}\log\frac{\rho +|\sigma |}{\rho -|\sigma |} -\frac{\rho}{2|\sigma
|^2}\right)\Psi _1\right\}\]}\\
\hskip-2em $\rho ^0$&$-1$\\
\hskip-2em $\sigma ^0$& $(\rho ^2-|\sigma |^2)^{-1}\sigma$
\end{tabular}
\end{ruledtabular}
\end{table*}

\subsection{Transformation to a Newman--Unti Frame}

The asymptotics of the tetrad and curvature components computed above
can all be examined as $s\to +\infty$, and compared with the
requirements for a Bondi--Sachs frame, as given for example in
ref.~\cite{PR1986}.  There are three sorts of adaptations which are
necessary to make the tetrad accord with the standard formulas.

One of these is to shift the zero of $s$ to eliminate the $s^{-2}$
term in the asymptotic expansion of $\rho$.  However, this only alters
subdominant terms in the expansions, and we will only need the
dominant terms, so we omit this.  The second change is to replace the
spinor $\iota ^A$ by one which becomes tangent to future null infinity
as $s\to\infty$; this can be accomplished by a null rotation.  The
final change is to break the local boost invariance of the dyad so as 
to
achieve the standard asymptotic scaling $\rho '\sim (2s)^{-1}$.  This
requires solving an elliptic equation on $\ess$ (equivalent to
conformally uniformizing the sphere).

Because of the additional computational resources required to solve
the elliptic equation on the sphere, we will distinguish here between
the two stages of the passage to the Bondi--Sachs frame.  When the
frame as been adjusted by a null rotation so one spinor is tangent to
null infinity, we call it a {\em Newman--Unti} frame and indicate it
by the postscript NU; after the rescaling, the Bondi--Sachs frame is
indicated by B.  Formulas valid in Newman--Unti frames will thus
automatically be valid in Bondi--Sachs frames.  We shall give the
transformation to a Newman--Unti frame in this section; the next
section will cover Bondi--Sachs frames.

The spinor $\iota ^A$ representing the ingoing null congruence has
been fixed by the geometry of the two-surface $\ess$, and propagated
outwards along null geodesics; there is no reason to expect it to be
asymptotically tangent to null infinity.  We must therefore anticipate
making a null rotation $\iota ^A\to \iota ^A_{\rm NU}=\iota
^A+Q\omicron ^A$ in order to achieve this.  To preserve the
parallel-propagation condition $D\iota ^A_{\rm NU}=0$, we shall need
$DQ=0$.  

The equation determining $Q$ comes from the fact that (with the 
correct choice
of $\iota ^A_{\rm NU}$) 
the quantity $\tau$ must vanish at least as $O(s^{-2})$ as $s\to 
\infty$
(cf.~\cite{NU1962,PR1986}). 
(Note that this means $Q$ will not transform homogeneously under
rescalings of the dyad.)
Since we have $\tau =\omicron ^A D'\omicron _A$, we find
$\tau _{\rm NU} = \omicron ^A(D+Q\delta
+\overline{Q} \delta ')\omicron _A=\tau +Q\sigma +\overline{Q} \rho$.  
Setting the lead asymptotic term of this to zero and using
the asymptotic forms of $\rho$, $\sigma$ and $\tau$, we see that
\begin{eqnarray}
Q&=& (\rho ^2-|\sigma |^2)^{-1}\Bigl\{ 
\left[\begin{array}{cc}\rho&\sigma \end{array}\right]
\left[\begin{array}{c}\overline\alpha +\beta\\
-\alpha-\overline\beta\end{array}\right] \nonumber\\
  &&-\sigma
  (\frac{\rho ^2-|\sigma |^2}{4|\sigma
|^3}\log\frac{\rho +|\sigma |}{\rho -|\sigma |} -\frac{\rho}{2|\sigma 
|^2}
)\overline{\Psi _1}\Bigr\}\Bigr| _\ess\, .
\end{eqnarray}

Now let us work out the operator $\delta$ for this frame, which we
denote $\delta _{\rm NU}=\delta +\overline{Q} D$.  We first invert the
definition of $\delta _\ess$ to get
\begin{eqnarray}
\left[\begin{array}{c}\delta\\ \delta '\end{array}\right]
  &=&(|A|^2 -|B|^2)^{-1}\left[\begin{array}{cc}A&-B\\ -\overline{B}
&\overline{A}\end{array}\right]
\left[\begin{array}{c}\delta _\ess -CD\\ \delta '_\ess -\overline{C}
D\end{array}\right] \nonumber\\
&\sim &  -s^{-1}(\rho _\ess ^2-|\sigma _\ess |^2)^{-1}\nonumber\\
&&\times  \left[\begin{array}{cc}\rho&-\sigma \\ -\overline{\sigma}
&\rho\end{array}\right]_\ess
\left[\begin{array}{c}\delta _\ess -CD\\ \delta '_\ess -\overline{C}
D\end{array}\right]
\end{eqnarray}
(where $\sim$ denotes asymptotic equality as $s\to +\infty$), so
\begin{eqnarray}
\left[\begin{array}{c}\delta _{\rm NU}\\ \delta '_{\rm NU}\end{array}
\right]
&\sim &-s^{-1}(\rho_\ess ^2-|\sigma_\ess |^2)^{-1}
  \left[\begin{array}{cc}\rho&-\sigma \\ -\overline{\sigma}
&\rho\end{array}\right]_\ess
\left[\begin{array}{c}\delta _\ess -CD\\ \delta '_\ess -\overline{C}
D\end{array}\right]\nonumber\\
 && +\left[\begin{array}{c}\overline{Q}\\ Q\end{array}\right] D\, .
\end{eqnarray}
Inserting the asymptotic values from Table~\ref{secondtable}, we find 
that the
coefficients of $D$ cancel, and we are left simply with
\begin{equation}\label{deltaeq}
\left[\begin{array}{c}\delta _{\rm NU}\\ \delta '_{\rm NU}\end{array}
\right]
\sim -s^{-1}(\rho _\ess ^2-|\sigma_\ess |^2)^{-1}
  \left[\begin{array}{cc}\rho&-\sigma \\ -\overline{\sigma}
&\rho\end{array}\right]_\ess
\left[\begin{array}{c}\delta _\ess \\ \delta '_\ess 
\end{array}\right]\, .
\end{equation}
This cancellation
is not a surprise; it arises from the fact that the $s=\const$ 
surfaces can
be taken be the ones approaching cuts of $\scrif$.

\begin{table*}
\caption{\label{thirdtable}Leading forms of the spin-coefficients and 
operators
in the Newman--Unti frame, under the assumption 
$|\sigma _\ess|\ll |\rho _\ess|$.
Each of these quantities falls off as $s^{-1}$ with the corresponding 
coefficient.  Thus $\alpha _{\rm NU}\sim s^{-1}\alpha ^0$, etc.  
In the right-hand column, all quantities are to be evaluated at $\ess
$.}
\begin{ruledtabular}
\begin{tabular}{cc}
Quantity&Value\\
$\delta _0$&$-\rho ^{-1}\delta _\ess$\\
$\delta' _0$&$-\rho ^{-1}\delta' _\ess$\\
$\alpha ^0$&$-\rho ^{-1}\{ -\overline\beta 
+(2\rho )^{-1}\overline{\Psi _1}\}$\\
$\beta ^0$&$-\rho ^{-1}\{ \beta -(2\rho )^{-1}\Psi _1\}$\\
$(\rho ')^0$&\parbox{50em}{\begin{eqnarray*}
 \rho ^{-1}\left\{\right.
  &-&\rho ' -2(3\rho ^2)^{-1} |\Psi _1|^2 
  +(2\rho ^2)^{-1} (\eth'\Psi _1+\eth\overline{\Psi _1}) 
  -(2\rho)^{-1}(\Psi _2+\overline{\Psi _2}) 
 -5(6\rho ^3)^{-1}(\Psi _1\eth'\rho +\overline{\Psi _1}\eth\rho )
  \\
   &-&
   (2\rho )^{-1}(\eth\overline\tau +\eth '\tau ) 
   +(2\rho ^2)^{-1}(\overline\tau \eth\rho +\tau\eth '\rho ) 
   +(2\rho ^2)^{-1}(\Psi _1\overline\tau +\overline{\Psi _1}\tau )
   +(2\rho ^2)^{-1}(\tau\eth\overline\sigma +\overline\tau 
\eth'\sigma ) \\
   &-&\left.
     (6\rho ^3)^{-1} (\Psi _1\eth\overline\sigma +\overline{\Psi _1} 
\eth '\sigma )\right\}
   \end{eqnarray*}}\\
$(\sigma ')^0$&   \parbox{50em}{\[
-\rho ^{-1}\left\{ \sigma '-\eth((2\rho ^2)^{-1}\overline{\Psi _1}-
\rho ^{-1}\tau )
  -\rho((2\rho ^2)^{-1}\overline{\Psi _1} -\rho ^{-1}\tau )^2\right\}
\]}
\end{tabular}
\end{ruledtabular}
\end{table*}

We now find the spin-coefficients $\alpha _{\rm NU}$, $\beta _{\rm NU}
$ 
in the Newman--Unti frame.  (We shall not use these explicitly, but we 
give them
both for completeness and for purposes of comparison with other work.)
The transformation rules for
these are
\begin{eqnarray}
\alpha _{\rm NU}&=&\alpha +Q\rho\\
\beta _{\rm NU}&=&\beta +Q\sigma\, .
\end{eqnarray}
Using the results from Table~\ref{secondtable}, we find
\begin{eqnarray}
\alpha _{\rm NU}\!\!\!&\sim &\!\!\!-s^{-1}
  (\rho _\ess ^2-|\sigma_\ess |^2)^{-1}\!\left\{
  -\rho\overline\beta +\overline\sigma \overline\alpha 
+(1/2)\overline{\Psi
_1}\right\}\Bigr|_\ess\qquad \\
\beta _{\rm NU}\!\!\!&\sim&\!\!\!-s^{-1}(\rho _\ess ^2-|\sigma_\ess |
^2)^{-1}\left\{ \rho
\beta -\sigma\alpha
-(1/2)\Psi _1\right\}\Bigr| _\ess\, ;
\end{eqnarray}
results in the approximation $|\sigma _\ess |\ll |\rho _\ess |$ are 
listed in 
Table~\ref{thirdtable}.

The fact that $Q$ remains bounded as $s\to\infty$ (in fact,
$Q$ is constant along the generators of $\N$), together with the
peeling property, means that transformation to the Newman--Unti frame 
does
not affect the asymptotic forms of $\Psi _1$, $\Psi _2$, $\Psi _3$.
Also $\rho _{\rm NU}=\rho$ and $\sigma _{\rm NU}=\sigma$, since these
only involve derivatives of $\omicron ^A$.

The transformations for the optical coefficients for the ingoing
congruence are more complex.  We have $\rho '_{\rm NU}=\rho '-\eth
Q-Q^2\sigma$, $\sigma '_{\rm NU}=\sigma '-\eth 'Q -Q^2\rho$.  We shall
not need these, but for completeness their asymptotic forms  are given
in Table~\ref{thirdtable}.

\section{Bondi--Sachs Frames}\label{BSF}

In this section, we complete the transformation to a Bondi--Sachs
frame, and also establish some of the calculus of these frames which
will be required for the analysis of energy--momentum and angular
momentum.  The Bondi--Sachs frames are essentially equivalent to the
notion of an ``asymptotic laboratory frame.''

From this point on, we will omit the subscript $\ess$ for
spin-coefficient quantities (including operators $\delta$, $\delta '$,
$\eth$, $\eth '$) at the two-surface, unless explicitly indicated
otherwise.  Quantities considered at other points on $\N$ will either
be in the Newman--Unti or the Bondi--Sachs frames and will be 
indicated by
subscripts NU or B.

\subsection{Transformation to a Bondi--Sachs Frame}

We recall that the Newman--Unti frame established in the previous
sections is very nearly a Bondi--Sachs frame; what remains is to
adjust the spin frame, or equivalently conformally transform, in order
to make the $s=\const$ cross-sections unit spheres.

Let us begin with the metric structure.  This is characterized by the
intrinsic $\delta$-operators of the $s=\const$ surfaces.
Equation~(\ref{deltaeq}) expressed these in terms of the structure at
$ \ess$; let us put $\delta _{\rm NU}=M^a\nabla _a$, so 
\begin{equation}
  M^a\sim -s^{-1}(\rho ^2-|\sigma |^2)^{-1}(\rho m^a-\sigma{\overline 
m}^a)\, .
\end{equation}
Then $-2M^{(a}{\overline M}^{b)}$ gives the intrinsic inverse metric
on the $s=\const$ surfaces.  The intrinsic metric is
$-2M_{(a}{\overline M}_{b)}$, where $M_a$, ${\overline M}_a$ are {\em
not} defined via lowering with $g_{ab}$, but via the duality relations
$M_a{\overline M}^a=-1$, $M_aM^a=0$; explicitly
\begin{equation}
  M_a\sim s(-\rho m_a-\sigma {\overline m}_a)
\end{equation}
and
\begin{equation}
-2M_{(a}{\overline M}_{b)}\sim -2s^2(\rho m_{(a}+\sigma {\overline 
m}_{(a})
  (\rho {\overline m}_{b)}+\overline\sigma m_{b)})\, .
\end{equation}

Under a change of scale $\omicron ^A\mapsto \lambdabar \omicron ^A$,
$\iota ^A\mapsto \lambdabar ^{-1}\iota ^A$, The factor $(\rho
m_{(a}+\sigma {\overline m}_{(a}) (\rho {\overline
m}_{b)}+\overline\sigma m_{b)})$ will be multiplied by $|\lambdabar |
^4$;
we may therefore regard the change of scale as equivalent to a
conformal transformation by $\Omega ^2=|\lambdabar |^4$.  More 
precisely,
for each choice of scale we have a family of metrics on the $s=\const$
surfaces which are scalar multiples of each other; when we change the
scale we also change the choices of the $s=\const$ surfaces (since
$s\mapsto |\lambdabar |^{-2}s$), and it is the metrics on these 
(pulled
back to $\ess$) which are conformally rescaled by $|\lambdabar |^4$
(relative the pull-back of the metric on the original surface at the
same numerical value of $s$).

The conformal structure is characterized by the complex structure.  We
may introduce a complex stereographic (antiholomorphic) coordinate
\footnote{The orientation on the sphere induced from its embedding in
space--time (or in a spacelike hypersurface with its induced
orientation) is opposite to the one used in ordinary complex analysis,
and so what is antiholomorphic from the space--time point of view is
holomorphic for ordinary complex analysis.} $\zeta$ on the $s=\const$
surfaces by requiring $M^a\nabla _a\zeta =0$, that is,
\begin{equation}\label{ahol}
  (\rho\delta  -\sigma \delta ' )\zeta =0
\end{equation}
and requiring that $\zeta$ be regular over $\ess$ except for a simple
pole (equivalently, that $ \zeta ^{-1}$ vanishes at a single point and
in the limit of approach to this point its argument has winding number
$-1$, the minus sign on account of its antiholomorphic character).  

The coordinate $\zeta$ is unique up to a fractional linear
transformation.  In order to keep its interpretation as direct as
possible, when $\ess$ is approximately a round sphere the coordinate
$\zeta$ should be taken to be close to a stereographic coordinate
$e^{i\phi}\cot (\theta /2) $ on $\ess$.  One way to fix the freedom
would be to require the pole to lie on the $+z$ coordinate axis, the
zero to lie on the $-z$ axis, and the point $\zeta =1$ to lie on the
$+x$ axis.  With these choices we effectively fix an asymptotic
``laboratory frame.''  (The time axis is fixed by the requirement that
$|\zeta |=1$ be a great circle.)

For numerical work, it may be more convenient to recast
Eq.~(\ref{ahol}) in terms of regular quantities.  If we let $\zeta _0$
be any smooth function on $\ess$ with a simple pole and simple zero of
the required type and write $\zeta = \hat\zeta \zeta _0$, then
$\hat\zeta$ is smooth over $\ess$ and satisfies the everywhere-regular
equation $M^a\nabla _a\hat \zeta =-\hat\zeta M^a\nabla _a\log\zeta
_0$.  There will be a one-complex-dimensional space of
everywhere-regular nowhere-vanishing solutions to the equation for
$\hat\zeta$.  These solutions will give $\zeta=\hat\zeta \zeta _0$ the
same pole and zero as $\zeta _0$; thus, if these have been chosen as
in the previous paragraph, one has simply to adjust the multiplicative
constant in $\hat\zeta$ to achieve the final normalization $\zeta =1$
on the $+x$ axis.  We shall assume from now on that a solution $\zeta$
to~(\ref{ahol}) has been found.

Now let us turn to the metric structure.  As is conventional, put
$M_a=-{\overline P}^{-1}d\zeta$, so that $-1={\overline M}^aM_a\sim 
{\overline P}^{-1}s^{-1}(\rho ^2-|\sigma |^2)^{-1}(\rho \delta' 
-\overline\sigma \delta  )\zeta \sim{\overline P}^{-1}s^{-1}\rho
^{-1}\delta '\zeta$ and
\begin{equation}\label{Peq}
  P\sim -s^{-1}\rho ^{-1}\delta \overline\zeta\, .
\end{equation}
Then the metric is $\sim -2|P|^{-2}d\zeta d\overline\zeta$.   

Now let us consider a change of scale to achieve a Bondi--Sachs frame.
Let this be $\omicron ^A \mapsto \omicron ^A_{\rm B}=\lambdabar 
\omicron
^A$.  We may keep $\zeta$ unchanged (it is a conformal invariant); we
have then $P\mapsto P_{\rm B} =-{s}_{\rm B}^{-1}{\rho _{\rm
B}}^{-1}\delta _{\rm B} \overline\zeta
\sim -{\overline\lambdabar}^{-2}s_{\rm B}^{-1}\rho ^{-1}\delta
\overline\zeta.$  We rescale the metric to a sphere of radius
$s_{\rm B}$ with $P_{\rm B} =2^{-1/2}s_{\rm B}^{-1}(1+|\zeta |^2)$.
This will align the time axis of the Bondi--Sachs system with that of
the laboratory frame, and give the standard spin frame adapted to the
coordinate $\zeta$.  We then find
\begin{equation}\label{lambdaeq}
  \lambdabar ^2=-2^{1/2}\rho ^{-1} (1+|\zeta |^2)^{-1}\delta '\zeta
\, .
\end{equation}
With $\lambdabar$ known, we may read off the value of any spin- and
boost-weighted quantity in the Bondi--Sachs frame from its value in a
Newman--Unti frame.  (Note that the original $\omicron ^A$, $\iota ^A$
may be {\em any} spin frame for which $m^a$ is tangent to $\ess$;
Eq.~(\ref{lambdaeq}) provides the correct transformation to the
frame adapted to $\zeta$.) 

In what follows, we shall need the shear in a Bondi--Sachs frame.  We
have, from Table~\ref{secondtable}, that $\sigma _{\rm NU}\sim
s^{-2}(\rho ^2-|\sigma |^2)^{-1}\sigma$.  Inserting the appropriate
rescalings, we find
\begin{equation}
\sigma _{\rm B}\sim s_{\rm B}^{-2}\lambdabar {\overline
\lambdabar}^{-3} 
   (\rho ^2-|\sigma |^2)^{-1}\sigma
\end{equation}
with $\lambdabar$ given by~(\ref{lambdaeq}).   In particular, the {\em
Bondi shear} is the coefficient of $s_{\rm B}^{-2}$, that is
\begin{equation}\label{Bshear}
\sigma ^0_{\rm B}=\lambdabar {\overline\lambdabar}^{-3} 
(\rho ^2-|\sigma |^2)^{-1}
  \sigma\, .
\end{equation}

Finally, we remark that an alternative (and somewhat more traditional)
route to fixing the conformal factor is to require that the Gaussian
curvature of the $s=\const$ surfaces becomes asymptotically constant.
Since the Gaussian curvature is $\sim -2\rho _{\rm NU} \rho '_{\rm
NU}$ and $\rho _{\rm NU}\sim -s^{-1}$, this leads to the requirement
that $\rho '_{\rm NU}\sim (2s)^{-1}$.  One can use
Table~\ref{thirdtable} and the transformation rules discussed at the
beginning of section~\ref{intsch} to write this as a second-order
partial differential equation for $|\lambdabar|$, which is equivalent 
to
the usual equation for finding a conformal transformation uniformizing
the Gaussian curvature.  The formulas for this are rather more
complicated than those given in the present subsection, however.

\subsection{The Angular Potential for the Shear}\label{angpotsubsec}

The Bondi shear, being a spin-weight two quantity, admits an angular
potential $\lambda _{\rm B}$ such that 
\begin{equation}\label{angpot} 
  s_{\rm B}^2\eth ^2_{\rm B}\lambda _{\rm B}=\sigma _{\rm B}^0
\end{equation} 
(or equivalently $\eth ^2_{\rm B}\lambda _{\rm B}=\sigma _{\rm B}$).
The use of this potential
facilitates the computation of the Bondi--Sachs energy, and the
potential also plays a central role in the analysis of angular
momentum.  The {\em electric} and {\em magnetic} parts of the shear
are $\sigma _{\rm el}=\eth ^2_{\rm B}\Re\lambda _{\rm B}$ and $\sigma 
_{\rm
mag}=i\eth ^2_{\rm B}\Im\lambda _{\rm B}$.

Equation~(\ref{angpot}) is easily solved when 
$P_{\rm B}=2^{-1/2}s_{\rm B}^{-1}(1+|\zeta |^2)$.  In
this case, it can be written as
\begin{equation}
 (1/2)\partial _{\overline\zeta} (1+\zeta\overline\zeta )^2\partial
_{\overline\zeta}\lambda _{\rm B}=\sigma _{\rm B}^0\, ,
\end{equation}
and a Green's function for the operator can easily be derived from the
relation $\partial _{\overline\zeta} {\overline\zeta}^{-1}=\pi
\delta ^{(2)}(\zeta ,\overline\zeta )$ (where the right-hand side is 
the usual
$\delta$-function in the $\zeta$-plane).  We find
\begin{equation}\label{lgf}
  \lambda _{\rm B}(\zeta,\overline\zeta ) =\int G(\zeta,
\overline\zeta ;\acute\zeta ,\acute{\overline\zeta}) \sigma _{\rm B}^0
(\acute\zeta ,\acute{\overline\zeta}) \, d\acute\zeta\wedge
d\acute{\overline\zeta} /(2i)\, ,
\end{equation}
where
\begin{equation}\label{gf}
G(\zeta,
\overline\zeta ;\acute\zeta ,\acute{\overline\zeta})=-\pi^{-1}
(\zeta -\acute\zeta )^{-1}\left(
\frac{\overline\zeta}{1+|\zeta|^2}
-\frac{\acute{\overline\zeta}}{1+|\acute\zeta |^2}\right)
\end{equation}
and 
\begin{equation}
(2i)^{-1}d\zeta\wedge d\overline\zeta 
  =(1-|\sigma /\rho |^2) |\delta\overline\zeta |^2 \, d
\ess\, .
\end{equation}

An alternative means of solving Eq.~(\ref{angpot}) would be to
resolve $\sigma _{\rm B}^0$ into spin-weighted spherical harmonics
${}_2Y_{j,m}$ and then use the relation $\eth _{\rm B}^2\,{}_0Y_{j,m}=
(1/2)\sqrt{(j-1)j(j+1)(j+2)}\,{}_2Y_{j,m}$ to infer the corresponding
resolution of $\lambda _{\rm B}$.  Thus one would have
\begin{eqnarray}
  \sigma ^0_{j,m}&=&2^{1/2}\int \overline{{}_2Y_{j,m}} (1+|\zeta|
^2)^{-1}
    \sqrt{\frac{(\delta '\zeta)^3}{(\delta \overline\zeta )}} 
      \rho ^{-1}\sigma \, d\ess\, ,\ \label{shearresa}\\
    \lambda _{j,m}&=&2((j-1)j(j+1)(j+2))^{-1/2}\sigma ^0_{j,m}
          \text{ (}j\geq 2\text{)}\, ,\qquad
\label{shearresb}\\
    \lambda _{\rm B}&=&\sum \lambda _{j,m}\, {}_0Y_{j,m}\, .
\label{shearresc}
\end{eqnarray}      
The terms $\lambda _{j,m}$ with $j=0,1$ are freely specifiable.  We
take these terms to vanish, which will simplify the coordinatization
of the twistor space, below.

The phases of the spin-weighted spherical harmonics depend on the
spin-frame.  There are two common choices:  that adapted to the
complex coordinate $\zeta$, and that adapted to $\theta$, $\phi$.
Because the analysis has been given here in terms of $\zeta$, it is
that spin-frame and those spherical harmonics which are used in
Eq.~(\ref{shearresa}).  To use the harmonics with respect to
$\theta$, $\phi$ one must, besides replacing ${}_{-2}Y_{j,m}(\zeta
,\overline\zeta )$ with ${}_{-2}Y_{j,m}(\theta ,\phi )$, also include
a factor of $(\zeta /\overline\zeta )^2$ in the integrand
of~(\ref{shearresa}).  (Cf. the appendix; see
ref.~\cite{PR1984} for a detailed discussion of the
harmonics.)

\section{Asymptotic Twistors and Spinors}

The Bondi--Sachs energy--momentum is a covector in a certain vector
space, the space of asymptotically constant covectors.  This space is
most easily constructed from the space of asymptotically constant
spinors.  Similarly, the twistorial angular momentum is defined on the
space of asymptotic twistors.  In fact, the asymptotic spinors are
naturally defined in terms of a canonical fibration of twistor space,
so we shall start by constructing the twistors and then specialize to
the spinors.  We conclude this section by giving the kinematic
twistor, in terms of which the energy--momentum and angular momentum
will be defined.

\subsection{The Twistor Space}

The twistor space $\T (\scrC (\ess ))$ of the cut $\scrC (\ess )$ of
null infinity associated with $ \ess$ is the set of solutions of the
two-surfaces twistor equation at $\scrC (\ess )$.  These equations are
\begin{eqnarray}
  \eth '_{\rm B}{\tilde\omega}^0_{\rm B}&=&0\label{twistoreqb}\\
  s\eth _{\rm B}\omega ^1_{\rm B}&=&\sigma _{\rm B}^0{\tilde
\omega}^0_{\rm B}\, ,
  \label{twistoreq}
\end{eqnarray}
where ${\tilde\omega}^0_{\rm B}=s^{-1}\omega ^0$ is rescaled to attain 
a finite limit
at $\scrC (\ess )$, and $s\eth _{\rm B}$ tends to an operator 
depending on angle
only; cf. Eq.~(\ref{Peq}).  There is a four-complex-dimensional space 
of solutions to
these which we shall give shortly.  For completeness, we note the 
forms of these
equations in terms of the spin-coefficients at $\ess$ are
\begin{eqnarray}
-(\rho\eth ' -\overline\sigma \eth +(1/2)\overline{\Psi _1}){\tilde
\omega}^0 
    &=&  0\\
-(\rho\eth -\sigma\eth ' +(1/2)\Psi _1)\omega ^1&=&\sigma {\tilde
\omega}^0
\, .
\end{eqnarray}
(The minus signs are included because $\rho$ is negative.)

\subsection{Solving the Twistor Equation}

Solutions to the twistor equation are easily found.  The
equation~(\ref{twistoreqb}) for ${\tilde \omega}^0$ has as its space 
of
solutions the spherical harmonics of spin-weight $-1/2$; thus
\begin{equation}\label{teeka}
  {\tilde\omega}^0_{\rm B}=
  2^{1/2}(1+|\zeta |^2)^{-1/2}(Z^3+Z^2\overline\zeta )\, ,
\end{equation}
where $Z^2$ and $Z^3$ are constants.  To solve the remaining equation,
we adopt a device of K.  P. Tod and set 
\begin{equation}\label{teekb}
\omega ^1_{\rm B}={\tilde\omega}^0_{\rm B}\eth _{\rm B}\lambda _{\rm 
B} 
    -\lambda _{\rm B}\eth _{\rm B}{\tilde\omega}^0_{\rm B} +\xi _{\rm 
B}\, ,
\end{equation}
where $\lambda _{\rm B}$ is an angular potential for the shear 
(subsection \ref{angpotsubsec}), 
and $\xi _{\rm B}$ is a spin-weight $+1/2$ quantity to be determined.  
We note that
\begin{equation}\label{teekc}
\eth _{\rm B}{\tilde\omega}^0_{\rm B}=(1+|\zeta|^2)^{-1/2}(Z^2-
Z^3\zeta )\, .
\end{equation}
Then the remaining equation~(\ref{twistoreq}) is equivalent to 
$\eth '_{\rm B}\xi _{\rm B}=0$, and the solutions to this are
\begin{equation}\label{teekd}
  \xi _{\rm B}=-i(1+|\zeta |^2)^{-1/2}( -Z^0+Z^1\zeta )\, .
\end{equation}
Thus $(Z^0,Z^1,Z^2,Z^3)$ coordinatize the twistor space; the factors
have been chosen to make them accord with those induced from the
standard Cartesian basis of Minkowski space if the cut of null
infinity is got from the light-cone of the origin (cf.~\cite{PR1984},
section 4.15 and \cite{PR1986}, section 6.1)\footnote{This differs
slightly from the basis common for work near null infinity; with our
$\omicron ^A$ being $2^{-1/4}$ of that one and our $\iota ^A$ being
$2^{1/4}$ of that one.}.

\subsection{Structures on Twistor Space}

There are three important structures on twistor space:  a fibration,
an infinity twistor, and a certain reality structure.  The fibration
and the infinity twistor allow the definition of asymptotic spinors;
the reality structure defines the null geodesics which play the roles
of origins for the definition of angular momentum.  Finally, the
reality structure and the infinity twistor combine to define a certain
twistor operation, the ``hook,'' which enters in the definition of
energy--momentum.

\subsubsection{The Fibration, the Infinity Twistor and Spinors}

The most primitive and important structure on twistor space is the
fibration, which is defined by simply keeping the ${\tilde\omega}^0$
field of the twistor.  This is just $(Z^0,Z^1,Z^2,Z^3)\mapsto
(Z^2,Z^3)$ in our coordinates, since ${\tilde \omega}^0$ is specified
by $Z^2$ and $Z^3$.  We see then that the space of fibers is a two-
complex-dimensional space; it is naturally identifiable with the space
of (dual, primed) {\em asymptotically constant spinors} $\Spin _{A'}$.
Asymptotic spinors of other valences and asymptotic vectors are
tensors are defined as usual by tensor operations from $\Spin _{A'}$.

Closely related to the fibration is the {\em infinity twistor}
$I(Z,\acute Z)=I_{\alpha\beta}Z^\alpha{\acute Z}^\beta$.  In our
coordinates, it is simply given by 
\begin{equation}
  I(Z,\acute Z) =Z^2{\acute Z}^3-Z^3{\acute Z}^2\, .
\end{equation}
This evidently defines a skew form on $\Spin _{A'}$, which represents
the asymptotically constant spinor $\epsilon ^{A'B'}$.  Its negative
inverse is $\epsilon _{A'B'}$; the spinor $\epsilon _{AB}\epsilon
_{A'B'}$ represents the asymptotic metric $g_{ab}=g_{AA'BB'}$. 
One often puts $Z^2=\pi _{0'}$,
$Z^3=\pi _{1'}$ and then one has
\begin{equation}
 I(Z,\acute Z )=\epsilon ^{A'B'}\pi _{A'}{\acute\pi}_{B'}
\end{equation}
with $\epsilon ^{0'1'}=1$ as usual. 

One can also define spin space $\Spin ^A$ directly in twistor terms,
as the kernel of the fibration; thus $\Spin ^A$ is identified with the
set of spinors whose coordinates are $(Z^0,Z^1,0,0)$.  However, it
will be more natural for us to work with $\Spin _{A'}$, especially
when we take up evolution.

\subsubsection{The Reality Structure}

The twistor space for Minkowski space is equipped with a sesquilinear
form of signature $+{}+{}-{}-$ whose zero set is the set of {\em real
twistors}.  In general relativity, there is also a reality structure,
but it is more nonlinear when spin is present.  The analytic
manifestation of this is that the candidate expression for the norm
\begin{equation}\label{tnorm}
i({\overline\omega}^{1'}_{\rm B}\eth _{\rm B}{\tilde\omega}^0_{\rm B}
-{\tilde\omega}^0_{\rm B}\eth _{\rm B}{\overline\omega}^{1'}_{\rm B}
+{\overline{\tilde\omega}}^{0'}_{\rm B}\eth '_{\rm B}\omega ^1_{\rm B}
-\omega ^1_{\rm B}\eth '_{\rm B}{\overline{\tilde\omega}}^{0'} _{\rm 
B})
\end{equation}
is not in general constant over $\ess$.  It turns out that we get a
good theory of angular momentum, however, by simply evaluating this
expression at the point on the sphere for which ${\tilde\omega}^0_{\rm 
B}$,
or equivalently ${\tilde\omega}^0$,
vanishes.  (There always will be
a unique such point, unless the field ${\tilde\omega}^0$ vanishes
identically, in which case the result is taken to be zero by a
continuity argument.)  We will denote the point at which
${\tilde\omega}^0$ vanishes by $\gamma ({\tilde\omega}^0)$; the
restriction of~(\ref{tnorm}) to $\gamma ({\tilde\omega}^0)$ by $\Phi
(Z)$.  A twistor is {\em real} iff $\Phi (Z)=0$.  The real twistors
correspond to real null geodesics meeting null infinity; they take
the place of space--time points as ``origins'' for the evaluation of
angular momentum.

Using the formulas above we find that
\begin{equation}
\Phi (Z)=\left[ -2\Im\lambda |\eth _{\rm B}{\tilde\omega}^0_{\rm B}|^2
  +i(\overline\xi\eth _{\rm B}{\tilde\omega}^0_{\rm B} 
    +\xi\eth '_{\rm B}{\overline{\tilde
\omega}}^{0'}_{\rm B})\right]\Bigr| _{\gamma ({\tilde\omega}^0)}\, .
\end{equation}
The stereographic coordinate of
$\gamma ({\tilde\omega}^0)$ is $\zeta = -\overline{Z^3/Z^2}$ and a 
brief calculation gives
\begin{equation}
\eth _{\rm B}{\tilde\omega}^0\Bigr| _{\gamma ({\tilde\omega}^0)}
  =\frac{Z^2}{|Z^2|}\sqrt{|Z^2|^2+|Z^3|^2}\, .
\end{equation}
Using these, we find
\begin{eqnarray}
\Phi (Z) &=&-2\Im\lambda \Bigr| _{\gamma ({\tilde\omega}^0)}
    (|Z^2|^2+|Z^3|^2)\nonumber\\
 &&+Z^0\overline{Z^2}+Z^1\overline{Z^3}+Z^2\overline{Z^0}
     +Z^3\overline{Z^1}\, .
\end{eqnarray}
Here the second line would be the usual twistor norm in special
relativity; the contribution on the first line is an essentially
general-relativistic effect.  We recall that $\Im\lambda$ is the
angular potential for the magnetic part of the shear; it is this
magnetic shear which distorts the twistor norm.  (The magnetic shear
is found to be the $j\geq 2$ components of the specific --- that is,
per unit mass --- spin.)

\subsubsection{The Hook Operation}

While, as discussed above, the special-relativistic twistor norm does
not extend to general relativity, enough of the structure does survive
that a certain antilinear operation does carry over to general
relativity.  This is the {\em hook} of a twistor $Z^\alpha$, denoted
by $I^{\beta\alpha}{\overline Z}_\beta$.  (Owing to the non-existence
of a norm, the quantity ${\overline Z}_\beta$ does not have any
separate meaning for us, but we shall keep the special-relativistic
notation.)  If we set ${\acute Z}^\alpha =I^{\beta\alpha}{\overline
Z}_\beta$, then the definition of the hook is
\begin{eqnarray}\label{hooka}
{\acute{\tilde\omega}}^0&=&0\\
{\acute{\tilde\omega}}^1&=&-i\overline{\tilde\omega}^{0'}\, .
\label{hookb}
\end{eqnarray} 
In our coordinates, the hook operation is $(Z^0,Z^1,Z^2,Z^3)\mapsto
(-\overline{Z^3},\overline{Z^2},0,0)$.  (The hook of any twistor is an
element of $\Spin ^A$.)

\subsection{The Kinematic Twistor}

In twistor theory, the energy--momentum and angular momentum are
encoded in a {\em kinematic twistor} $A(Z)=A_{\alpha\beta}Z^\alpha
Z^\beta$, defined by
\begin{equation}
\begin{split}
A(Z) = -i(4\pi G)^{-1}
\oint\left\{ \right.&\Psi _1^{\rm NU} (\omega ^0_{\rm NU})^2
  +2\Psi _2^{\rm NU}\omega ^0_{\rm NU}\omega ^1_{\rm NU} \\
  +&\left.\Psi _3^{\rm NU}
(\omega ^1_{\rm NU})^2\right\} \, dS_{\rm NU}\, ,
\end{split}
\end{equation}
where the limit as the surface tends to null infinity is understood.  
The kinematic twistor satisfies the Hermiticity property
$A_{\alpha\beta}{\acute Z}^\alpha I^{\gamma\beta}{\overline Z}_\gamma 
=
\overline{A_{\alpha\beta}Z^\alpha 
I^{\gamma\beta}{\acute{\overline{Z}}}_\gamma}$,
which in our coordinates is
\begin{eqnarray}
  A_{00}=A_{01}&=&A_{11}=0\label{herma}\\
  A_{02}=-\overline{A_{13}}\, ,\ A_{12}&=&\overline{A_{12}}\, ,\ 
  A_{03}=\overline{A_{03}}\, .\label{hermb}
\end{eqnarray}

To work out the kinematic twistor explicitly, let us first insert
the asymptotic forms of the quantities; we find
\begin{equation}
\begin{split}
A(Z)=&-i(4\pi G)\oint (\rho ^2 -|\sigma |^2) \\
&\times\left\{
  \Psi _1^0 ({\tilde\omega}^0)^2 +2\Psi _2^0{\tilde\omega}^0\omega ^1
  +\Psi _3^0 (\omega ^1)^2\right\}\,  d\ess\, ,
\end{split}
\end{equation}
in terms of data on $\ess$. 

We next express the twistors in terms of their forms in the
Bondi--Sachs frame, taking into account the rescaling relative to the
Newman--Unti one (recall that ${\tilde\omega}^0$, $\omega ^1$ have
Newman--Penrose types $\{ 0,1\}$, $\{ 1,0\}$ respectively):
\begin{equation}\label{kteq}
\begin{split}
A(Z)&=-i(4\pi G)
\oint\left\{
  \Psi _1^0 {\overline\lambdabar}^{-2}
       ({\tilde\omega}^0 _{\rm B})^2 \right.\\
  &\left.+2\Psi _2^0
(\lambdabar\overline\lambdabar )^{-1}
      {\tilde\omega}^0_{\rm B}\omega ^1_{\rm B}
  +\Psi _3^0 \lambdabar ^{-2}(\omega ^1_{\rm B})^2\right\}\, 
       (\rho ^2 -|\sigma |^2)
   \, d\ess\, .
\end{split}
\end{equation}

At this point, we may substitute the explicit forms of the solutions
of the twistor equation given in this section to compute the
components $A_{\alpha\beta}$ of the kinematic twistor.  Because of
round-off errors (and also the approximation $|\sigma |\ll {|\rho |}$,
if used), the
numerical computation of $A_{\alpha\beta}$ directly from~(\ref{kteq})
could fail to satisfy the Hermiticity conditions (\ref{herma}),
(\ref{hermb}); we therefore enforce these conditions at the levels of
the integrands.  The results of this are given in
Table~\ref{fourthtable}.  

\begin{table*}
\caption{\label{fourthtable}  Integrands for the computation of the
potentially non-zero
components $A_{\alpha\beta}=A_{\beta\alpha}$ of the kinematic twistor.
Each term on the right is to be multiplied by the common factor $-i
(4\pi G)^{-1} (1+|\zeta |^2)^{-1}(\rho ^2 -|\sigma |^2)$ and 
integrated with
respect to $d\ess$ over the sphere.  
In this table, the Hermiticity conditions $A_{12}= \overline{A_{12}}$, 
$A_{03}=\overline{A_{03}}$, $A_{02}=-\overline{A_{13}}$ have been 
enforced at the level of the integrands. }
\begin{ruledtabular}
\begin{tabular}{cc}
Component&Quantity to
be multiplied by $-i (4\pi G)^{-1} (1+|\zeta |^2)^{-1}(\rho ^2-|\sigma
|^2)$ and integrated with 
respect to $d\ess$\\
$A_{02}=-\overline{A_{13}}$
  &$2^{1/2}i\overline\zeta |\lambdabar |^{-2}\Re\Psi _2^0
  +i\overline{\zeta} \Re (\Psi _3^0\lambdabar ^{-2} 
         2^{1/2}\eth _{\rm B}\lambda _{\rm B})
+2^{-1}i({\overline\zeta}^2-1)
        \Re (\Psi _3^0\lambdabar ^{-2}\lambda _{\rm B})
  -2^{-1}({\overline\zeta}^2+1)
         \Im (\Psi _3^0\lambdabar ^{-2}\lambda _{\rm B})$\\
$A_{03}$
 &$2^{1/2}i|\lambdabar |^{-2}\Re\Psi _2^0
   +i\Re (\Psi _3^0\lambdabar ^{-2}(2^{1/2}\eth _{\rm B}\lambda _{\rm 
B} 
           +\zeta\lambda 
_{\rm B}))$\\
$A_{12}$
 &$-2^{1/2}i|\zeta |^2 |\lambdabar |^{-2}\Re\Psi _2^0
   -i\Re (\Psi _3^0\zeta (2^{1/2}\overline\zeta \eth _{\rm B}\lambda 
_{\rm B}
          -\lambda _{\rm 
B}))$\\
$A_{22}$
 &$2\Psi _1^0{\overline\lambdabar}^{-2}{\overline\zeta}^2 
         +2^{3/2}\Psi _2^0 |\lambdabar |
^{-2} {\overline\zeta} (2^{1/2}\overline\zeta \eth _{\rm B}\lambda 
_{\rm B}
          -\lambda _{\rm B}) 
+\Psi _3^0\lambdabar ^{-2} 
     (2^{1/2}\overline\zeta \eth _{\rm B}\lambdabar _{\rm B} -
           \lambda _{\rm B})^2$\\
$A_{23}$
 &$2\Psi _1^0{\overline\lambdabar}^{-2}\overline\zeta
   +2^{1/2}\Psi _2^0 |\lambdabar |^{-2}
             (2^{3/2}\overline\zeta\eth _{\rm B}\lambda _{\rm B} 
+(|\zeta |^2-1)\lambda _{\rm B}) 
     +\Psi _3^0\lambdabar ^{-2}
                   (2^{1/2}\eth _{\rm B}\lambda _{\rm B} +\zeta\lambda 
_{\rm  B})    
        (2^{1/2}\overline\zeta \eth _{\rm B}\lambda _{\rm B}-\lambda 
_{\rm B})$\\
$A_{33}$
 &$2\Psi _1^0{\overline\lambdabar}^{-2}+2^{3/2}\Psi _2^0 |
                     \lambdabar |^{-2}(2^{1/2}\eth _{\rm B}\lambda 
_{\rm B}
                              +\zeta\lambda _{\rm B})  
          +\Psi _3^0\lambdabar ^{-2}(2^{1/2}\eth _{\rm B}\lambda _{\rm 
B} 
                        +\zeta\lambda _{\rm B})^2$
\end{tabular}
\end{ruledtabular}
\end{table*}

\section{Energy--Momentum}

The Bondi--Sachs energy--momentum $P_a$ is most naturally viewed as a
function $P^{AA'}{\overline\pi}_A\pi _{A'}$ on the space $\Spin _{A'}$
of asymptotically constant spinors.  In twistor terms, this is
\begin{equation}\label{enmom}
P^{AA'}{\overline\pi}_A\pi _{A'} =A_{\alpha\beta} Z^\alpha I^{\gamma
\beta} 
{\overline Z}_\gamma\, ,
\end{equation}
where the right-hand side represents the contraction of the kinematic
twistor once with $Z^\alpha$ and once with its hook
$I^{\gamma\beta}{\overline Z}_\gamma$; the result~(\ref{enmom}) is
real depends only on the projection $\pi _{A'}$ of the twistor $Z$;
that is, if the twistor components are $(Z^0,Z^1,Z^2=\pi _{0'},Z^3=\pi
_{1'})$, the choice of $Z^0$ and $Z^1$ is immaterial.

Using the formula for the hook map in coordinates
(just below Eq.~(\ref{hookb})), we find explicitly
\begin{eqnarray}\label{Benmom}
\left[\begin{array}{cc} P^{00'}&P^{01'}\\ P^{10'}&P^{11'}\end{array}
\right]
  &=&2^{-1/2}\left[\begin{array}{cc} P^t+P^z&P^x+iP^y\\ P^x-iP^y&P^t-
P^z
      \end{array}\right]\\
      &=&\left[\begin{array}{cc} A_{12}&A_{13}
    \\ -A_{02}&-A_{03}
    \end{array}\right]
\, .\nonumber
\end{eqnarray}

\section{Angular Momentum}

In special relativity, the angular momentum is a position-dependent
skew tensor field $M_{ab}(x)$, or equivalently a spinor $\mu
_{A'B'}(x)$, where $M_{ab}(x)=\mu _{A'B'}(x)\epsilon
_{AB}+{\overline\mu}_{AB}(x)\epsilon _{A'B'}$.  In general relativity,
as is well known, there is no good set of ``asymptotic origins'' for
the measurement of angular momentum, and thus the special-relativistic
treatment does not apply.

A good treatment of angular momentum in general relativity is
possible, if we adjust our perspective a bit.  We first note that if
$\geod$ is a null geodesic in Minkowski space and $\pi _{A'}$ is a
tangent spinor to it, then the component $\mu^{A'B'}(x)\pi _{A'}\pi
_{B'}$ of the angular momentum is (by the change-of-origin formula)
independent of $x$, as long as $x$ lies on $\geod$.  Thus we may think
of the angular momentum either as a spinor-valued function on
space--time, or as $\mu (\geod ,\pi _{A'}) =\mu ^{A'B'}(x)\pi _{A'}\pi
_{B'}$, a function on the space of null geodesics together with their
tangent spinors.  While angular momentum does not extend to general
relativity as a spinor-valued function, it does extend as a function
of the null geodesic and the tangent spinor.  Indeed, the expression 
is 
very simple:  we have
\begin{equation}
 \mu (\geod ,\pi _{A'})=(2i)^{-1}A_{\alpha\beta}Z^\alpha Z^\beta\, ,
\end{equation}
where $Z\leftrightarrow (\geod ,\pi _{A'})$ is the real twistor 
defined by 
the null geodesic $\geod$ and the tangent spinor $\pi _{A'}$.  

While the general-relativistic angular momentum is thus an extension
of the special-relativistic concept, some features which are prominent
on the special-relativistic case do not extend to general relativity,
and others which are usually viewed as secondary become central in the
general-relativistic setting.

The root of this is that the general-relativistic angular momentum is
defined on the space $\{ (\geod ,\pi _{A'})\}$ of null geodesics
together with their tangent spinors.   In the asymptotic
regime, this space naturally fibers over the space $\Spin _{A'}$ of
spinors, because there is a well-defined asymptotic spin space.  This
contrasts with the usual view of $\mu ^{A'B'}(x)\pi _{A'}\pi _{B'}$
being defined on the spin bundle $\{ (x,\pi _{A'})\}$ of Minkowski
space, where the {\em base} space is Minkowski space and the {\em
fibers} are copies of $\Spin _{A'}$.  In practical terms, this means
that while the component of the angular momentum $\mu (\geod ,\pi
_{A'})$ in a direction corresponding to $\geod$ and $\pi _{A'}$ will
be well-defined, there will be no natural way of simultaneously
varying $\geod$ and $\pi _{A'}$ so the angular momentum is specified
by a pure $j=1$ representation $\mu ^{A'B'}$ of the Lorentz group; the
angular momentum will inevitably (if  a magnetic part of the shear is
present) have $j\geq 2$ parts as well.

The reason for the appearance of these $j\geq 2$ components is that
general relativity unifies the ``ordinary'' ($j=1$) angular momentum
with gravitational radiation.  The $j\geq 2$ parts of the angular
momentum correspond exactly to the shear (times the Bondi mass).
Because there is no split of the angular momentum into $j=1$ and
$j\geq 2$ parts with the appropriate geometric invariance (invariance
under the Bondi--Metzner--Sachs group), one must, to get an invariant
theory, consider all of the $j\geq 1$ parts of the angular momentum.

While the angular momentum does depend on the pairs $(\geod, \pi
_{A'})$, the dependence is not arbitrary:  there is a
general-relativistic analog of the change-of-origin formula, which
says that the angular momenta at different points in a fiber differ by
appropriate multiples of the components of the energy--momentum.  Thus
the full information in the angular momentum can be recovered by
choosing any cross-section $\geod (\pi _{A'})$ of the fibration and
computing $\mu (\geod (\pi _{A'}), \pi _{A'})$ as $\pi _{A'}$ varies.
Because this is a homogeneous function (of degree two) in $\pi _{A'}$,
the essential information in the angular momentum is that in one
spin-weight minus one function on the sphere.

It is natural for us to choose the cross-section to be given by the
congruence of null geodesics meeting the cut of null infinity
orthogonally (that is, the congruence specified by $l^a$).  This
congruence then serves as a sort of origin for the computation.
(However, the present prescription differs essentially from previous
attempts to use cuts as origins.)  Then the electric and magnetic
parts of $\mu (\geod (\pi _{A'}),\pi _{A'})$ represent the energy
moments and spatial angular momentum, respectively, of the system with
respect to the asymptotic laboratory frame.

We will also want a general-relativistic extension of the 
``polarized'' 
form $\mu ^{A'B'}\pi _{A'}
{\acute\pi}_{B'}$.  This corresponds to a two-point function 
\begin{equation}
\mu ((\geod ,\pi _{A'}),(\acute\geod ,{\acute\pi}_{A'}))
  =(2i)^{-1}A_{\alpha\beta}Z^\alpha {\acute Z}^\beta
\end{equation}
on twistor space, where $Z\leftrightarrow (\geod ,\pi _{A'})$, 
$\acute Z\leftrightarrow (\acute
\geod ,{\acute\pi}_{A'})$.  In special relativity we would have 
$(2i)^{-1}A(Z,\acute Z ) =\mu  ^{A'B'}(x_{\rm av})\pi _{A'}{\acute
\pi}_{B'}$, 
where $x_{\rm av}$ is any point on the world-line 
defined by ``averaging'' the geodesics $\geod$, $\acute\geod$ with 
respect to the energy--momentum ($x_{\rm av}=(x+\acute x )/2$ where 
$x\in\geod$, $\acute x\in\acute\geod$ satisfy 
$(x^a-{\acute x}^a)P_a=0$)~\cite{ADH2007}.  While in general 
relativity 
there is no similar invariant notion of averaging null geodesics, that 
is a 
limitation only on interpreting the origin of  the angular momentum in 
direct space--time terms and not on its well-definition as a conserved 
quantity.  

If we fix a cross-section of twistor space, then the angular momentum
 $\mu ((\geod (\pi _{A'}),\pi
_{A'}),(\geod ({\acute\pi}_{A'}),{\acute\pi}_{A'}))$ can be thought of 
as 
a two-point function on the sphere.  However, the essential 
information 
in it corresponds to functions of one point on the sphere, not two.  
This is 
because the condition (\ref{herma}) implies that the higher-$j$ terms 
enter 
only in tensor products with $j=1/2$, $s=-1/2$ representations, that 
is, 
the angular momentum is in fact the symmetrized tensor product of a 
single spin-weight minus one-half function with an ordinary spinor. 

The intrinsic spin may be computed by passing to a boosted frame in
which the time-axis is aligned with the Bondi--Sachs energy--momentum;
then the magnetic part of the angular momentum is the spin.  (The
electric part of the angular momentum in this frame has a natural
interpretation, too.  The twistorial construction makes the cut appear
as if it were a supertranslated cut in a stationary space--time; the
electric part is the Bondi mass times this supertranslation.)

\subsection{Reporting the Angular Momentum}

Besides the need to accommodate $j\geq 1$ representations, there is 
another issue to address in choosing how to report the angular 
momentum, 
which is a trade-off between invariance and intuitive familiarity.  
This 
issue already occurs in special relativity, where the angular momentum 
is 
invariantly an element of the $j=1$, $s=-1$ representation, but when a 
reference frame is chosen we usually think of it as two spatial 
vectors 
(really the elements of a complex $j=1$, $s=0$ representation).  We 
shall 
opt for the familiar presentation, so that we can speak of the spatial 
angular momentum and energy-moment (both with contributions for 
$j\geq 1$) parts of the full angular momentum.

Consider for a moment special relativity.  Let $t^a$ be a unit
future-pointing timelike vector, and let $z^a$ be a unit spacelike
vector orthogonal to it. (For this paragraph only it will be
convenient to regard $z^a$ as a variable direction on the sphere.)
Then $2^{-1/2}(t^a+z^a)$ is a future-directed null vector, say
${\overline\pi}^A\pi ^{A'}$ (normalized by $t_{AA'}{\overline\pi}^A\pi
^{A'}=2^{-1/2}$), and $t_A{}^{(A'}\pi ^{B')}{\overline\pi}^A
=t^{A(A'}z^{B')B}\epsilon _{AB}$.  It follows that
\begin{eqnarray}
  2^{1/2}\mu _{A'B'}t_A{}^{A'}{\overline\pi}^A\pi ^{B'}
    &=&M_{ab}t^{A(A'}z^{B')B}\nonumber\\
    &=&M_{ab}(1/2)(t^az^b -(i/2)\epsilon ^{abcd}t_cz_d)\nonumber\\
    &=&(1/2)(M_{tz}-iM_{xy})\, ,
\end{eqnarray}
where $(x,y,z)$ form a right-handed spatial triad.  Thus, having fixed
$t^a$, as $\pi _{A'}$ varies, one gets the energy--moment $M_{tz}$ and
spatial angular momentum $M_{xy}$ in the direction
$z^{AA'}=2^{1/2}{\overline\pi}^A\pi ^{A'}-t^{AA'}$ it determines.

We shall do the same thing in general relativity.  We take for $t^a$
the time direction determined by the asymptotic laboratory frame, we
allow $Z^2=\pi _{0'}$, $Z^3=\pi _{1'}$ to vary (normalized to
$|Z^2|^2+|Z^3|^2=1$), and we take
${\acute\pi}_{A'}=2^{1/2}t_{AA'}{\overline\pi}^A$, that is
${\acute\pi}_{0'}={\overline\pi}_1$,
${\acute\pi}_{1'}=-{\overline\pi}_0$.  With these restrictions $\mu
(\pi _{A'}) =\mu ((\geod (\pi _{A'}),\pi _{A'}),(\geod
({\acute\pi}_{A'}),{\acute\pi}_{A'}))$ becomes a spin-weight zero
function on the sphere, with $\mu +\overline\mu$ giving the
energy--moment, and $i\mu -i\overline\mu$ giving the spatial angular
momentum, in the direction determined by $\pi _{A'}$ (and $t^a$).
These may be reported as real functions on the sphere, or resolved
into spherical harmonics.  

(Of course, there is some freedom in choosing how to extend the 
terminology 
appropriate to a purely $j=1$ quantity to a $j\geq 1$ one.  For 
instance, 
what one chooses to call the 
$j\geq 2$ energy--moments and spatial angular momenta
could be taken to be some function of $j$ times the ones used here.
Such differences are unimportant here, since what we are 
interested in is simply extracting the invariant information.)

\subsection{Derivation of the Formula}

The twistorial formula for the angular momentum is simply
\begin{equation}\label{angmomeq}
\mu =(2i)^{-1}A_{\alpha\beta}Z^\alpha{\acute Z}^\beta\, ,
\end{equation}
where $A_{\alpha\beta}$ is the kinematic twistor introduced earlier
and $Z^\alpha$, ${\acute Z}^\alpha$ are twistors whose null geodesics
meet the cut of null infinity orthogonally (and satisfy
$|Z^2|^2+|Z^3|^2=1$, ${\acute Z}^2=\overline{Z^3}$, ${\acute
Z}^3=-\overline{Z^2}$).  The condition that $Z^\alpha$ (say) meet the
cut at a point is that the fields $({\tilde\omega}^0_{\rm B},\omega
^1_{\rm B})$ vanish there; that the meeting be orthogonal means that
the tangent spinor to the geodesic must lie in the $\omicron _{A'}$
direction.  

Let the point in question on the cut have stereographic coordinate
$\zeta$.  Then from the formulas~(\ref{teeka}), (\ref{teekb}),
(\ref{teekc}), (\ref{teekd}) for the twistor fields, we deduce that
the conditions for $\geod$ to meet the cut are
\begin{eqnarray}
  \zeta &=&-\overline{Z^3/Z^2}\, ,\label{ztval}\\
  \lambda _{\rm B}(|Z^2|^2+|Z^3|^2) &=& 
    i(Z^0\overline{Z^2}+Z^1\overline{Z^3})\, .
\end{eqnarray}
The component of the tangent spinor in the $\iota ^{A'}_{\rm B}$
direction, which we require to vanish, was computed in Eq.~(29) of
ref.~\cite{ADH2007}.  It is
\begin{equation}
\begin{split}
  (|Z^2|^2+|Z^3|^2)^{-1/2} &|Z^2|^{-1}\times\\
  (i (\eth '_{\rm B}\lambda _{\rm B} )
     (|Z^2|^2+|Z^3|^2) Z^2 &+2^{-1/2}(Z^0Z^3-Z^1Z^2) 
\overline{Z^2} )\, ,
 \end{split}
\end{equation}
where $\eth '_{\rm B}\lambda _{\rm B}$ is evaluated at~(\ref{ztval}).

After a little algebra, we find the equations for $Z^0$, $Z^1$ in
terms of $Z^2=\pi _{0'}$, $Z^3=\pi _{1'}$:
\begin{eqnarray} 
  Z^0&=&-i(2^{1/2}Z^2{\overline{Z^3}}\eth '_{\rm B}\lambda _{\rm B} 
  +|Z^2|^2\lambda _{\rm B})/\overline{Z^2}\, ,\label{Z0eq}\\
  Z^1&=&i(2^{1/2}Z^2\eth '_{\rm B}\lambda _{\rm B} -Z^3\lambda _{\rm 
B})\, ,
  \label{Z1eq}
\end{eqnarray}
where $\lambda _{\rm B}$ and $\eth '_{\rm B}\lambda _{\rm
B}=2^{-1/2}(1+|\zeta |^2)\partial\lambda _{\rm B}/\partial\zeta$ are
evaluated at $\zeta$.  Eqs.~(\ref{Z0eq}), (\ref{Z1eq}) determine the 
cross-section of twistor space as a bundle over spin space.

The same analysis applies, of course, with $Z^\alpha$ replaced by
${\acute Z}^\alpha$.  We note that $\acute\zeta =-1/\overline\zeta$ is
the point antipodal to $\zeta$ on the sphere, and the quantities
$\lambda _{\rm B}$, $\eth _{\rm B}\lambda _{\rm B}$ appearing in the
formulas for ${\acute Z}^0$, ${\acute Z}^1$ must be evaluated at this
antipodal point.

Substituting these formulas into Eq.~(\ref{angmomeq}) and collecting 
like 
terms, we find (with the normalization $|Z^2|^2+|Z^3|^2=1$) that
\begin{widetext}
\begin{equation}\label{angmomeqzeta}
\begin{split}
\mu =2^{-1}(1+|\zeta |^2)^{-1}\left\{\right.
  &-[A_{13}+(A_{12}+A_{03})\zeta +A_{02}\zeta ^2]
         2^{1/2}\eth '_{\rm B}\lambda _{\rm B} 
(\zeta ,\overline\zeta )
    +[A_{03}+A_{02}\zeta -A_{13}\overline\zeta -A_{12}|\zeta |^2]
        \lambda _{\rm B}(\zeta ,
\overline\zeta )\\
    &+[A_{02}-(A_{03}+A_{12})\overline\zeta +A_{13}{\overline\zeta}^2]
         (\zeta /\overline
\zeta ) 2^{1/2}\eth '_{\rm B}\lambda _{\rm B}(-1/\overline
\zeta ,-1/\zeta )\\
    &\qquad
    +[A_{12}+A_{02}\zeta-A_{13}\overline\zeta -A_{03}|\zeta |^2]
          \lambda _{\rm B}
(-1/\overline\zeta ,-1/\zeta )\\
    &\left.+iA_{22}\zeta -iA_{23}(|\zeta |^2-1) -iA_{33}\overline\zeta
\right\}\, .
\end{split}
\end{equation}
\end{widetext}
(Here we have followed the standard convention of writing a non-
holomorphic 
function $f$ of  the complex coordinate $\zeta$ as $f(\zeta ,\overline
\zeta )$.)
While the formula is lengthy, this is mostly due to the appropriate 
factors of 
$\zeta$, $\overline\zeta$ for weighting the components $A_{\alpha
\beta}$ 
of the kinematic twistor; the only complicated expressions are those 
involving $\lambda _{\rm B}$, which encode the shear.

Thus Eq.~(\ref{angmomeqzeta}) allows one to report the angular 
momentum as 
a function of the direction (with $2\Re\mu$ giving the energy--moment 
in the 
direction, and $-2\Im\mu$ the spatial angular momentum about the axis, 
specified by $\zeta$).  

If the resolution of $\lambda _{\rm B}$ in spherical harmonics is 
known, 
one can use it to find the resolution of $\mu$ into spherical 
harmonics, 
by identifying the explicit functions of $\zeta$ in 
(\ref{angmomeqzeta}) with 
particular spin-weighted harmonics and applying tensor product 
formulas 
(``Clebsch--Gordan decompositions'').  The computation is lengthy but 
straightforward using the formulas derived in the appendix; we find 
\begin{equation}\label{angmomsh}
\begin{split}
&\mu =\sum {\hat\mu}_{j,m}\, {}_0Y_{j,m}\\ 
 &+(2i)^{-1}\sqrt{2\pi /3}(A_{22}\, {}_0Y_{1,1}+2^{1/2}A_{23}\, 
{}_0Y_{1,0}+A_{33}\, 
{}_0Y_{1,-1})
\, ,
\end{split}
\end{equation}
where the last three-terms are the $\lambda _{\rm B}$-independent 
ones, and
the coefficient ${\hat\mu}_{j,m}$ is a sum of terms, each of which is 
$\lambda _{j'm'}$ for $j'=j-1,j,j+1$, $m'=m-1,m,m+1$ times a factor; 
these are 
given in Table~\ref{fifthtable} (for $j$ even) and Table~
\ref{sixthtable} (for $j$ odd).

\begin{table*}[!]
\caption{\label{fifthtable} Terms contributing to the 
$\lambda _{\rm B}$-dependent part of the 
angular momentum proportional to ${}_0Y_{jm}$
for {\em even} $j$.  Each term is the product of the component $
\lambda 
_{j',m'}$ in the left column by the quantity in the right column.  
(We understand the term is zero 
unless $j'=0,1,2,\ldots$, $m'=-j',\ldots ,j'$.)}
\begin{ruledtabular}
\begin{tabular}{cc}
Component $\lambda _{j',m'}$&Factor it multiplies\\
$\lambda _{j,m-1}$
  &$2^{-1}A_{02}((j-m+1)(j+m))^{1/2}$\\
$\lambda _{j,m}$
 &$-2^{-1}(A_{03}+A_{12})m$\\
$\lambda _{j,m+1}$
  &$-2^{-1}A_{13}((j+m+1)(j-m))^{1/2}$\\
\end{tabular}
\end{ruledtabular}
\end{table*}

\begin{table*}[!]
\caption{\label{sixthtable} Terms contributing to the $\lambda _{\rm 
B}$-dependent part of the 
angular momentum proportional to ${}_0Y_{jm}$
for {\em odd} $j$.  Each term is the product of the component $\lambda 
_{j',m'}$ in the left column by the quantity in the right column.  (We 
understand the term is zero 
unless $j'=0,1,2,\ldots$, $m'=-j',\ldots ,j'$.)}
\begin{ruledtabular}
\begin{tabular}{cc}
Component $\lambda _{j',m'}$&Factor it multiplies\\
$\lambda _{j-1,m-1}$
 &$2^{-1}A_{02}(j-2)((j+m)(j-m+1)/(4j^2-1))^{1/2}$\\
$\lambda _{j+1,m-1}$
 &$2^{-1}A_{02}(j+1)((j-m+2)(j-m+1)/(4(j+1)^2-1))^{1/2}$\\
$\lambda _{j,m}$
 &$2^{-1}(A_{03}-A_{12}) $\\
$\lambda _{j-1,m+1}$
 &$2^{-1}A_{13}(j-2)((j-m)(j-m-1)/(4j^2-1))^{1/2}$\\
$\lambda _{j+1,m+1}$
  &$2^{-1}A_{13}(j+1)((j+m+2)(j+m+1)/(4(j+1)^2-1))^{1/2}$
\end{tabular}
\end{ruledtabular}
\end{table*}

\section{Evolution}\label{EV}

The usefulness of a definition of energy--momentum or angular momentum
in radiation problems depends considerably on whether it admits a
well-defined notion of evolution.  At null infinity, it is well-known
that the Bondi--Sachs energy--momenta at two cuts can be compared.
Many proposed definitions of angular momentum at null
infinity took values in cut-dependent spaces, making tracking
their evolution problematic; the twistor-based definition solves this
problem.  Here, however, because we are dealing with energy--momentum
and angular momentum at large but {\em finite} spheres $\ess$, we
must take up the problem anew, both for energy--momentum and angular
momentum.

Suppose we have a one-parameter family of surfaces $\ess (\eta )$ (for 
$\eta$ in some interval $J$) foliating a timelike surface $T$, with 
$\eta$ 
increasing towards the future (precisely, we require $v^a\nabla _a\eta 
>0$ for every future-causal vector $v^a$ tangent to $T$).  We may 
compute the 
energy--momentum and angular momentum on each of these; the difficulty 
is that these quantities are naturally defined on the null infinities 
$\scrif (\eta )$ of the different auxiliary space--times $M(\eta )$ 
(each defined by taking $\Psi _0=0$ along the null hypersurface 
$\N (\eta )$ outwards 
from $\ess (\eta )$).  In order to compare the energy--momenta and 
angular momenta for different $\eta$, then, we must find a natural way 
of identifying the null infinities $\scrif (\eta )$ for different $
\eta$.

We could express this as a problem of finding transition functions.  
The constructions above determine a preferred Bondi coordinate system 
$(\zeta _\eta ,\overline{\zeta _\eta},u_\eta )$ on $\scrif (\eta )$.
(We have shown how to fix, for each $\ess (\eta )$, a complex 
stereographic coordinate $\zeta =\zeta _\eta$ on $\scrif (\eta )$, and 
we have chosen an associated Bondi--Sachs frame by fixing the factor 
$P_{\rm B}$.  This determines the Bondi retarded time $u=u_\eta$ up to 
a supertranslation; we fix this by requiring the preferred cut 
$\scrC (\eta )$ of $\scrif (\eta )$ --- the limit of $\N (\eta )$ --- 
have $u_\eta =0$.)  
Then we wish to find formulas for $(\zeta _{\eta _2},\overline{\zeta 
_{\eta _2}}, u_{\eta _2})$ in terms of 
$(\zeta _{\eta _1},\overline{\zeta _{\eta _1}}, u_{\eta _1})$.

\subsubsection*{Structures Preserved by the Identifications}

I will discuss the way these identifications are made 
shortly.  More important, though, is the question,
What structures are preserved by the
identifications?  Were the surfaces $\ess (\eta )$ actually cuts
of the null infinity of the physical space--time, we should expect the 
usual structures to be preserved and the identifications to be 
Bondi--Metzner--Sachs transformations.  However, here we must expect a 
weaker structure due to finite-size effects.  The extent to which 
these effects are significant should be interpreted as the extent to 
which the extraction surfaces $\ess (\eta )$ are insufficiently 
distant to capture the full radiative structure (or, if the effects 
persists for arbitrarily large surfaces, the extent to which the 
Bondi--Sachs asymptotics fail for the physical space--time).  Even 
with this interpretation, though, in order to precisely quantify the 
finite-size effects we must work out the structure in general.

The usual intrinsic structure of null infinity may be regarded as 
determined by three elements:  its set of generators, each with an 
affine structure; a conformal 
structure on the set of generators 
(it is a remarkable feature of the construction 
that the set does have a well-defined conformal structure); and the 
``strong conformal geometry'' (which links the scales of vectors up 
the generators with the scales of the area forms transverse to them).  
In our case, two of these three elements survive:  there are natural 
invariant definitions of the generators and of the strong conformal 
geometry, but the conformal structure on the space of generators is 
{\em not} preserved under the identifications.

{\em A priori,} while for each value of $\eta$ each generator of $
\scrif (\eta )$ has an affine structure, it is not evident that there 
is a preferred way of identifying these for different values of $\eta
$.  However,
because the strong conformal geometry and the space of generators are 
well-defined,
the identifications will extend in a natural way to vectors tangent to 
the generators.  This means that supertranslations are well-defined.  
We shall see that there is a natural way of measuring the 
supertranslation relating $u_{\eta _1}=0$ to $u _{\eta _2}=0$, and 
this will allow us to appropriately account for the change of 
section of the twistor space when we compare angular momentum.

The failure of an invariant conformal structure to exist on the space 
of generators  (that is, to 
be preserved under evolution) turns out to mean that that in comparing 
the energy--momenta at different cuts higher-$j$ representations 
appear.  

To see this, let us first recall that, since the space of generators 
has naturally the smooth structure of an oriented sphere, a conformal 
structure on it is equivalent to a complex structure.  In the Bondi--
Sachs case, the transition functions $(\zeta _{\eta _1},
\overline{\zeta _{\eta _1}})\mapsto (\zeta _{\eta _2},\overline{\zeta 
_{\eta _2}})$ preserve this complex structure, and are thus fractional 
linear transformations.  
On the other hand, these fractional linear transformations are 
isomorphic to the (proper, isochronous) Lorentz transformations.  Thus 
in the Bondi--Sachs case, the admissible coordinate changes induce 
Lorentz transformations on the space of generators.  One builds up the 
spaces of asymptotically constant spinors, vectors, etc., as functions 
on the space of generators, and it is the fact that the conformal 
transformations induce Lorentz motions which is responsible for these 
fields breaking neatly into Lorentz-invariant representations.

Now let us turn to the present, non-Bondi--Sachs case.  If we compute 
the component
$P({\eta _1};\zeta 
_{\eta _1},\overline{\zeta _{\eta _1}})$ of the energy--momentum
at $\eta _1$ along the null vector determined by the Bondi 
stereographic coordinates $(\zeta _{\eta _1},\overline{\zeta _{\eta 
_1}})$ in the chart at $\eta _1$, we find, as usual, that, as a 
function of $(\zeta _{\eta _1},\overline{\zeta _{\eta 
_1}})$, the energy--momentum consists of $j=0$ and $j=1$ components, 
forming a covector.
However, if we want to compare the energy--momenta at $\eta _1$ and $
\eta _2$, we must express them both in a common chart, say the chart $
(\zeta _{\eta _2},{\overline\zeta}_{\eta _2})$ on 
the space of generators for $\scrif (\eta _2)$.  
Then of course $P({\eta _2};\zeta 
_{\eta _2},\overline{\zeta _{\eta _2}})$ will have only $j=0$ and $j=1$ 
components, but,
because the change of variables
$(\zeta _{\eta _1},\overline{\zeta _{\eta _1}})\mapsto
(\zeta _{\eta _2},\overline{\zeta _{\eta _2}})$ will not be a 
fractional linear transformation, the expression for $P(\eta _1)$ in 
terms of
$(\zeta _{\eta _2},{\overline\zeta}_{\eta _2})$ will
generally contain not just $j=0$ and $j=1$ components, but those for 
all integral $j$.
So comparison of energy--momenta at different surfaces $\ess (\eta 
_1)$, $\ess (\eta _2)$, will require higher-$j$ representations.
(This sort of behavior 
occurs even strictly at null infinity for angular momentum, but is a 
finite-size effect for energy--momentum.)

Again, 
this potential failure of a linear identification of asymptotic 
covectors (and elements of the spin-tensor algebra generally)
as $\eta$ changes is a finite-size effect; it will become 
negligible if one takes the family $\ess (\eta )$ of extraction 
surfaces distant enough (that is, close enough to the physical 
space--time's null infinity), assuming that the system is indeed 
isolated.  Thus the occurrence of these 
nonlinearities in a numerical computation would be a signal that the 
extraction surface had not been taken large enough that a model null 
infinity, with the usual Bondi--Sachs structure, stable under 
evolution existed.

Some further, more technical, discussion of structure is given in
Subsection~\ref{Tech}.

\subsubsection*{How the Identifications are Made}

A few words now
about how the identifications of $\scrif (\eta )$ for 
different $\eta$ are made.  They all grow out of 
two considerations, which we have already used extensively.  The first 
of these is that at any point $p(\eta )\in \ess (\eta )$, there is a 
null geodesic orthogonally outwards in $M(\eta )$ whose end-point lies 
on $\scrif (\eta )$; holding $\eta$ fixed but varying $p(\eta )$ we 
get a preferred cut $\scrC (\eta )$of $\scrif (\eta )$.  The second is 
that 
at any $p(\eta )\in\ess (\eta )$, there is a canonical 
isomorphism $T_p(M_{\rm phys})\cong T_p(M(\eta ))$ between the tangent 
space of the physical space--time and that of the auxiliary space 
$M(\eta )$.  
(Cf. footnote~[\theefn].)

We define the identifications at the infinitesimal level (that is, for 
infinitesimally separated $\eta$), and then integrate.  At the 
infinitesimal level, the problem comes down to understanding how the 
the structures on $\scrif (\eta +d\eta )$ are represented by 
quantities on $\scrif (\eta )$.

Let us consider a one-parameter family of points $p(\eta )\in \ess 
(\eta )$ (with $\eta$ the parameter) in $M_{\rm phys}$.  
At each point we have 
the null normal $l^a$ orthogonally outwards form $\ess (\eta )$, and 
its associated null geodesic in $M_{\rm phys}$.  As $\eta$ varies, the 
vector $u^a_{\rm phys}$ connecting this family of geodesics is a 
Jacobi 
field, which in turn is determined by its initial data $(u^a_{\rm 
phys},{\dot u}^b_{\rm phys})$ at $\ess (\eta )$.  On the other hand, 
we may use the isomorphism of tangent spaces to regard these as 
initial data for a Jacobi field $u^a$ in $M(\eta )$.  
We will take the
limiting 
value of {\em this} field at $\scrif (\eta )$ as the {\em definition} of the 
rate of change of the end-point of the null geodesic orthogonally 
outwards from $p(\eta )$ as $\eta$ varies.
The 
important thing to note here is that this vector, while it lies in the 
tangent space to $\scrif (\eta )$, represents the 
end-point of a geodesic with base-point $p(\eta +d\eta )$ for an 
infinitesimally differing value of $\eta$.  This is the root of all of 
the identifications.

We may apply this in several ways.  If, for example, we require the 
curve $p(\eta )$ to be such that the corresponding vector at $\scrif 
(\eta )$ points along a generator, we may say the generator of null 
infinity does not change with $p(\eta )$; this gives the 
identification of the generators of null infinity for different values 
of $\eta$.  Or if we imagine a congruence of curves, say $p(\zeta
,\overline\zeta ,\eta )$ with stereographic coordinate $\zeta$, we get
a vector at each point of the cut of $\scrif (\eta )$ labeled by
$\zeta$, and this vector field over the cut gives
a measure the supertranslation induced by changing $\eta$
infinitesimally, that is, in passing from $\ess (\eta )$ to $\ess
(\eta +d\eta )$.  

\subsubsection*{Outline of this Section}

Subsection~\ref{JacF} gives the computation of the Jacobi fields, and
Subsection~\ref{CompN} the main formulas for comparing the null
infinities.  Then Subsections~\ref{CEM},\ref{CAM} give the formulas 
for
treating the evolution of the energy--momentum and angular momentum.
The final subsection discusses some technical aspects of the structure
of the null infinities; these are not needed for the computations in
this paper but are given for completeness of the conceptual framework.

\subsubsection*{Coordinates}

In what follows, we will be comparing structure on the timelike 
hypersurface $T$ with that on $\scrif (\eta )$, and also structure on 
$\scrif (\eta _1)$ and $\scrif (\eta _2)$.
As a ready reference, here is a summary of the coordinate systems to 
be used.

Recall that, for each $\eta$, we have already defined coordinates
$(\zeta ,\overline\zeta )=(\zeta _\eta ,{\overline{\zeta _\eta}})$ on 
$\ess (\eta )$.  We may regard $(\zeta ,\overline\zeta )$ then as 
defined over the whole of $T$.

We shall eventually use coordinates $(\eta ,\zeta,\overline\zeta )$ on 
$T$.  However, in the next subsection it will be convenient to briefly 
use coordinates $(\eta ,\xi ,\overline\xi )$ where $\xi$ need not be 
simply related to $\zeta$.

As already indicated, we will have Bondi
coordinates $(\zeta _\eta ,\overline{\zeta _\eta} ,u_\eta )$ on $
\scrif (\eta )$.

\subsection{The Jacobi Fields}\label{JacF}

We have a family of spacelike surfaces $\ess (\eta )$ forming a 
timelike hypersurface $T$, with $v^a\nabla _a\eta >0$ for any 
future-pointing vector $v^a$ tangent to $T$.  
In practice, it is convenient to represent the evolution from one 
surface to the next by a connecting vector field $w^a$, that is, a 
field tangent to $T$ with $w^a\nabla _a\eta =1$.  The freedom in 
choosing $w^a$ is the freedom to add a vector field which, at each $
\eta$, is tangent to $\ess (\eta )$, that is, is a linear combination 
of $m^a$ and ${\overline m^a}$.  We shall see below that there is a 
natural way to fix this freedom, but for now we leave it unspecified.

It will be helpful to briefly use coordinates adapted to the foliation 
of $T$ by $\eta$ and the integral curves of $w^a$.  Near any point of 
interest on $T$, fix
$\eta _0$ and let $(\xi ,\overline\xi )$ be a complex coordinatization 
of $\ess (\eta _0)$.
(The use of a complex coordinate is only for brevity of treatment; the 
coordinate $\xi$ need not be a holomorphic function of $\zeta$, or 
have any other special relation to it.)
We may let 
the integral curve of $w^a$ with 
coordinates $(\xi ,\overline\xi )$ at $\eta _0$
be $p(\eta ,\xi ,\overline\xi )$.  If we Lie-drag  
$\xi$ along $w^a$ (so $w^a\nabla _a\xi =0$), then $(\eta ,\xi ,
\overline\xi )$ provides a coordinatization of a portion of $T$.  In 
these coordinates, we have $w^a=\partial /\partial \eta$, and $
\partial /\partial\xi $ is a linear combination of $m^a$ and $
{\overline m}^a$.

\subsubsection{Definition of the Fields; their Initial Data}

In this subsubsection, we work in the physical space--time.  to avoid cumbersome notation, however, the Jacobi field is denoted simply 
$u^a$, rather than $u^a_{\rm phys}$.

Along each integral curve $p(\eta ,\xi ,\overline\xi )$ of $w^a$ 
(holding $\xi$ fixed), let $\gamma (\eta ,s,\xi ,\overline\xi )$ be 
the affinely parameterized null geodesic outwards 
from $\ess (\eta )$ in $M_{\rm phys}$, so that $l^a=\partial _s\gamma 
(\eta ,s,\xi ,\overline\xi )$
and $\gamma (\eta ,0,\xi ,\overline\xi )=p (\eta ,\xi ,\overline\xi )
$.  (The scale of the affine parameterization will not matter.)  Then 
differentiating along $\eta$ we get a Jacobi field $u^a=\partial _\eta 
\gamma (\eta ,s,\xi ,\overline\xi )$ connecting these geodesics.  

It is this Jacobi field $u^a$ we wish to work out; more precisely, we 
wish to work out $u^a$ modulo terms proportional to $l^a$.  The field 
is determined by its initial data.
One of those data is simply $u^a\Bigr| _{s=0}=w^a$; the other is 
$l^b\nabla _b u^a\Bigr| _{s=0}$.  One constraint on this second datum 
is that $l_al^b\nabla _bu^a=0$.  (This follows by differentiating 
$l_al^a=0$:  we have $0=u^b\nabla _b(l^al_a)=2l_au^b\nabla _bl^a =2l_a 
l^b\nabla _bl^a$.)  The other constraint (affecting one complex degree 
of freedom) comes from requiring that the geodesics meet $\ess (\eta )
$ orthogonally.

The condition that the geodesics $\gamma (\eta ,s,\xi ,\overline\xi )$ 
meet $\ess (\eta )$ orthogonally is that $(\partial _\xi p)^al_a=0$.  
If we differentiate this along $w^a$, and we apply the conditions $w^b
\nabla _b(\partial _\xi p)^a=(\partial _\xi p)^b\nabla _b w^a$ (which 
holds since $w^a=\partial /\partial \eta $ in the $(\eta ,\xi ,
\overline\xi )$ coordinates on $T$) and $w^b\nabla _bl^a=u^b\nabla _b 
l^a=l^b\nabla _bu^a$ (which holds because $u^a$ is a connecting vector 
for the geodesics with tangents $l^a$), we find
\begin{equation}
  ((\partial _\xi p)^b\nabla _b w^a)l_a 
    +(\partial _\xi p)_a l^b\nabla _b u^a=0\, .
\end{equation}
However, since $(\partial _\xi p)^b$ is a complex basis vector 
spanning the tangent space to $\ess (\eta )$, and so the equation is 
equivalent to one with this vector
replaced by any other complex basis vector for this tangent space; in 
particular, it is equivalent to
\begin{equation}
  (m^b\nabla _b w^a)l_a 
    +m_a l^b\nabla _b u^a=0\, ,
\end{equation}
which determines the complex datum $m_a l^b\nabla _b u^a\Bigr| 
_{s=0}$.

We shall not need the coordinate $\xi$ in what follows.

\subsubsection{Solving the Jacobi Equation}

We have derived the initial data for the Jacobi fields on the physical 
space--time $M_{\rm phys}$.  We now make use of the isomorphism of 
tangent spaces at $\ess (\eta )$
to regard these same data as determining Jacobi fields 
in the auxiliary space--time $M(\eta )$, and solve the Jacobi equation 
there.  (Since this isomorphism is $u^a_{\rm phys}\mapsto u^a$ and we have already dropped the ``phys'' subscript, this amounts to simply using the formulas derived above for their data at $\ess (\eta )$.) 
Ultimately we are interested in the vectors on 
$\scrif (\eta )$ determined by the asymptotic forms of the Jacobi fields.

In this subsubsection, we work in a frame parallel-propagated along the 
null geodesics, and express the Jacobi fields in terms of their 
initial data.  In the next one, we will transform the asymptotic form 
of the Jacobi
field to the Bondi frame.  

It will be convenient to put
\begin{equation} 
  u^a=u^{00'}l^a +u^{01'}m^a+u^{10'}{\overline m}^a +u^{11'}n^a\, .
\end{equation}
Then the constraints we have worked out above will be the initial 
conditions 
for the geodesic deviation equation:
\begin{equation}\label{lasteq}
 u^a\Bigr| _{s=0}=w^a\text{  and  }
   l^a\nabla _a u^{10'}\Bigr| _{s=0}=l^bm^a\nabla _aw_b\, ;
\end{equation}
recall that $w^a$ is the vector field connecting $\ess (\eta )$ to $
\ess (\eta +d\eta )$, 
so the quantities on the right-hand sides in Eq.~(\ref{lasteq}) are 
known.  As noted above,
that the Jacobi field represent a null geodesic entails additionally
\begin{equation}
 l^a\nabla _a u^{11'}=0\, .
\end{equation}
The geodesic deviation equation itself 
($l^a\nabla _al^b\nabla _b u^c=l^pl^q R_{prq}{}^cu^r$)
becomes in terms of the components
\begin{eqnarray}
{\ddot u}^{00'}&=& \Psi _1u^{01'}+\Psi _2 u^{11'} +\text{conjugate}
\\
{\ddot u}^{10'}&=&-\Psi _1 u^{11'}
\\
{\ddot u}^{11'}&=&0
\, ,
\end{eqnarray} 
where the dots are differentiation with respect to $s$.
(Here of course the components of $u^a$ and also the curvatures 
$\Psi _1$, $\Psi _2$ are evaluated at points along the geodesic, 
not at $\ess =\ess (\eta )$; we temporarily violate the convention 
that quantities 
unsubscripted by NU or B are evaluated at $\ess$.)  Integrating these 
with the initial conditions (using the explicit form of $\Psi_1$ 
provided by 
integrating (\ref{Psi1eq}) using (\ref{optsol})), we find
\begin{equation}
 u^{11'}(s)= w^{11'}
\end{equation}
 and
\begin{equation}
\begin{split}
 u^{10'}(s)&=w^{10'}+sl^bm^a\nabla _aw_b+
  w^{11'}\Psi _{1} (4|\sigma |^2)^{-1}\times\\
  \{ 
   &-\log (1-s(\rho +|\sigma |))(1-s(\rho -|\sigma |) -2\rho s\\
  & -|\sigma |^{-1}(\rho -|\sigma |)(1-s(\rho +|\sigma |)\log (-s(\rho 
+|\sigma |) 
   \\
   &
     +|\sigma |^{-1}(\rho +|\sigma |)(1-s(\rho -|\sigma |)\log (-
s(\rho -|\sigma |) 
   \}\, ,
\end{split}
\end{equation}
where, on the right-hand side, the spin-coefficients and the curvature 
$\Psi _1$ 
(as well as the field $w^a$) are evaluated at $\ess (\eta)$ (we 
restore the 
convention about 
subscripting).  
We note the asymptotic behavior
\begin{equation}\label{geneq}
u^{10'}\sim s(l^bm^a\nabla _aw_b+w^{11'}X)\, ,
\end{equation}
where
\begin{equation}\label{Xeq}
X=\Psi _{1} \Bigl( -2^{-1}|\sigma |^{-2}\rho 
     +4^{-1}|\sigma |^{-3} (\rho ^2-|\sigma |^2)
     \log\frac{\rho +|\sigma |}{\rho -|\sigma |}\Bigr)\, .
\end{equation}
(In the limit $|\sigma |\ll |\rho|$, we 
have
$X\approx -(3\rho)^{-1}\Psi _{1}$.)

We will not need $u^{00'}$.

\subsubsection{Transformation to the Bondi Frame}

In the previous subsubsection, we found the Jacobi field (modulo terms 
tangent to the geodesic) in a parallel-propagated frame.  We here 
transform the asymptotic form of this field to the Bondi frame 
previously constructed for $M (\eta )$.  

We have 
\begin{equation}\label{ueq}
u^a\mbox{ modulo } l^a=u^{01'}m^a+u^{10'}{\overline m}^a +u^{11'} n^a
 \, .
\end{equation}
Now $m^a$ differs from $m^a_{\rm NU}$ by a term proportional to $l^a$, 
and so we may replace $m^a$ by $m^a_{\rm NU}$ in this expression.  As 
$s\to \infty$, we may also, to leading order, replace $n^a$ by 
$n^a_{\rm NU}$.  To see this, first note that we
have seen that $u^{01'},u^{10'}=O(s)$, $u^{11'}=O(1)$ as $s\to\infty$.  
On the other hand, we have 
$n^a=n^a_{\rm NU}-Qm^a-{\overline Q}{\overline m}^a +Q{\overline Q}l^a
$.  Since $Q$ is $O(1)$, making this substitution in Eq.~(\ref{ueq}) 
would only change the coefficients of $m^a$, ${\overline m}^a$ (or 
$m^a_{\rm NU}$, ${\overline m}^a_{\rm NU}$) by subdominant terms.  
Thus 
\begin{eqnarray}\label{uas}
u^a\mbox{ modulo } l^a&\sim& u^{01'}m^a_{\rm NU}+u^{10'}{\overline 
m}^a_{\rm NU} +u^{11'} n^a_{\rm NU}\nonumber\\
 &\sim&u^{01'}P\frac{\partial}{\partial{\overline\zeta}_\eta} +
  u^{10'}{\overline P}\frac{\partial}{\partial{\zeta _\eta}} +
  u^{11'}n^a_{\rm NU}\nonumber\\
  &\sim&
  -(l^b{\overline m}^a\nabla _aw_b+w^{11'}{\overline X})\rho ^{-1}
(\delta \overline\zeta)
\frac{\partial}{\partial{\overline\zeta}_\eta} \nonumber\\ 
  &&-(l^bm^a\nabla _aw_b+w^{11'}X)\rho ^{-1}(\delta '\zeta)
\frac{\partial}{\partial\zeta _\eta} \nonumber\\ 
  &&+w^{11'}|\lambdabar |^2\frac{\partial}{\partial u_\eta}
 \, .
\end{eqnarray}
Here $\partial /\partial u_\eta =n^a_{\rm B}$, where $u_\eta$ is a 
Bondi retarded time coordinate for $\scrif (\eta )$ adapted to the 
frame defined by $(\zeta _\eta,{\overline\zeta}_\eta )$ (and $P_{\rm 
B}$).

Equation~(\ref{uas}) represents the displacement, in $\scrif 
(\eta )$, of the cut formed from the null vectors orthogonally 
outwards, as the two-surface moves
from $\ess (\eta )$ to $\ess (\eta +d\eta )$ along the vector field 
$w^a$.  It thus codes the relation between the null infinities $\scrif 
(\eta )$ and $\scrif (\eta +d\eta )$; our next task is to develop this 
into formulas for transition functions.

\subsection{Comparison of Null Infinities}\label{CompN}

We now have the tools to compare the null infinities $\scrif (\eta )$ 
associated with different values of $\eta$.  We recall that (for each 
$\eta$) the invariant structures of $\scrif (\eta )$ are its space of 
generators, the conformal structure on that space, and the ``strong 
conformal geometry;'' we shall see that the first and last of these 
can be identified under changes of $\eta$, but not the conformal 
structure on the space of generators.

\subsubsection{Identification of the Space of Generators}

We have a family of two-surfaces $\ess (\eta )$, and for each of these 
the null geodesics outwards determine a cut of the corresponding null 
infinity $\scrif (\eta )$.  We saw in the last subsection, however, 
that we could represent the cut of an $\ess (\eta +d\eta )$ 
infinitesimally 
perturbed from $\ess (\eta )$ by a vector field in $\scrif (\eta )$.  
Precisely, if $w^a$ was a vector field at $\ess (\eta )$ with $w^a
\nabla _a\eta =1$, then Eq.~(\ref{uas}) gave the corresponding 
apparent displacement of the cut.

We may use this to identify the generators of $\scrif (\eta )$ for 
different values of $\eta$.  The connecting field $w^a$ will preserve 
the generator if the corresponding field at $\scrif (\eta )$ points 
purely up the generator, which, from Eq.~(\ref{uas}), is evidently if
\begin{equation}
l^bm^a\nabla _aw_b+w^{11'}X =0\, .
\end{equation}
Expanding $l^bm^a\nabla _aw_b$ 
in 
spin-coefficients we find
\begin{equation}\label{hceq}
   \eth w^{11'}-\sigma w^{01'}-\rho w^{10'}
  =w^{11'}X\, .
\end{equation} 
Here $w^{11'}=w^al_a$ depends only on the displacement of $\ess (\eta 
+d\eta )$ 
relative to $\ess (\eta )$; it is insensitive to the horizontal 
components  of $w^a$, 
which are $w^{01'}$ and its conjugate.  We may thus use (\ref{hceq}) 
as an equation 
to determine the horizontal components of $w^a$ from the condition 
that 
(\ref{geneq}) vanishes.  After a little algebra, we find
\begin{equation}\label{weq}
  w^{10'}=(\rho ^2-|\sigma |^2)^{-1}
  \left[\begin{array}{cc} \rho &-\sigma\end{array}\right]
  \left[\begin{array}{c} \eth w^{11'}+Xw^{11'}\\ 
    \eth' w^{11'}+{\overline X} w^{11'}\end{array}\right]\, .
\end{equation}

We may therefore determine a vector field $w^a$ on $T$ by 
requiring $w^a\nabla _a\eta =1$ and
its components tangential to $\ess (\eta )$ to be given by 
Eq.~(\ref{weq}).  Each integral curve of this vector field corresponds 
to a generator of null infinity, in the sense that under the 
identification of the $\scrif (\eta )$ for different $\eta$'s 
described here, the null geodesics orthogonally outwards from $\ess 
(\eta )$ along this curve are all considered to strike the same 
generator. 

An equivalent way of expressing this is in terms of transition 
functions for the angular coordinates.  Let us write $\zeta _{\eta 
_0}$ for the stereographic coordinate on $\scrif (\eta _0)$ determined 
by restricting the stereographic coordinate $\zeta$ to $\ess (\eta 
_0)$ (and identifying the cut of $\scrif (\eta _0 )$ with $\ess (\eta 
_0)$ by using the ideal end-points of the null geodesics orthogonally 
outwards).  We may then extend $\zeta _{\eta _0}$ to $\scrif (\eta )$ 
for all $\eta$ by requiring $w^a\nabla _a\zeta _{\eta _0} =0$.  

Now let $\cz (\zeta _0,\overline{\zeta _0},\eta _0,\eta )$ be the 
value of $\zeta =\zeta _\eta$ at $\ess (\eta )$ corresponding to the 
same generator as does the value $\zeta _0$ at $\ess (\eta _0)$.  Then 
$\cz (\zeta _0,\overline{\zeta _0},\eta _0,\eta )=\zeta _\eta\circ 
\zeta _{\eta _0}^{-1}$ can be regarded as the transition function from 
$\zeta _{\eta _0}$ to $\zeta _\eta$, with $\eta$, $\eta _0$ 
parameterizing the particular choices of coordinate function of 
interest.  

The derivative of $\cz$ with respect to $\eta$ will be the component 
$w^\zeta$ of $w^a$, in the coordinates $(\eta ,\zeta ,\overline\zeta )
$, if $w^a$ is chosen to preserve the generators of null infinity.  
It will be conceptually useful to put $w^a=w^a_{\rm v}+w^a_{\rm h}$, 
where $w^a_{\rm v}=w^{00'}l^a+w^{11'}n^a$ are the ``vertical'' 
components and $w^a_{\rm h}=w^{01'}m^a+w^{10'}{\overline m}^a$ are the 
``horizontal'' components (with respect to the foliation of $T$ by $
\eta$ and the induced metric).  Then $w^\zeta =w^a\nabla _a\zeta =w^
\zeta _{\rm v}+w^\zeta _{\rm h}$ where $w^\zeta _{\rm v,h}=w^a_{\rm 
v,h}\nabla _a\zeta$.  We have
\begin{equation}
\begin{split}
 w^a_{\rm h}&= w^{10'}{\overline m}^a+\text{conjugate}\\
 & =-w^{10'}s(\rho M^a+\overline\sigma {\overline M}^a) +
\text{conjugate}\\
  &=-w^{10'}s(\rho P\frac{\partial}{\partial\zeta} 
  +\overline\sigma \overline{P} \frac{\partial}
{\partial\overline\zeta} )+\text{conjugate}\\
  &=-s(w^{10'}\rho  +w^{01'}\sigma )P\frac{\partial}{\partial\zeta}
  +\text{conjugate}\\
  &=-(w^{10'}\rho  
  +w^{01'}\sigma )\rho ^{-1}\delta\overline\zeta \frac{\partial}
{\partial \zeta}+\text{conjugate}\, ,
\end{split}
\end{equation}  
%%
%% There was a sign error in the previous draft for this equation
%% 
and so $w^\zeta _{\rm h}=-(w^{10'}\rho +w^{01'}\sigma ) \rho 
^{-1}\delta
\overline\zeta$.  Using Eq.~(\ref{weq}), this becomes
$w^\zeta _{\rm h}=-(\eta w^{11'} +Xw^{11'})\rho ^{-1}\delta\overline
\zeta$, 
and so the transition function $\cz$ is determined by
\begin{subequations}\label{czeq}
\begin{eqnarray}
\frac{d\cz}{d\eta} &=&-(\eth w^{11'}+Xw^{11'}) \rho ^{-1}\delta
\overline\zeta +w^\zeta _{\rm v}\qquad\label{dcza}
  \\
  \cz (\zeta _0 ,\overline{\zeta _0} ,\eta _0,\eta _0) &=&\zeta _0\, .
  \end{eqnarray}
\end{subequations}
%%
%% There was a sign error in the previous draft for the first equation
%% also the vertical term was omitted
(This has been written as a family of ordinary differential equations 
parameterized by $\zeta _0$, $\overline{\zeta _0}$, $\eta$ to 
emphasize that it is a simple evolution equation; one could write $
\partial /\partial\eta$ in place of $d/d\eta$ just as correctly.)

In Eqn.~(\ref{dcza}), the two terms on the right represent 
contributions which are, respectively, intrinsic and extrinsic to $
\ess (\eta )$ in their dependence on the coordinate $\zeta$.  The 
first term can be thought of as representing the rate of deformation 
of the coordinate $\zeta$ due to changes in the geometry of the 
surface as $\eta$ is increased; the second may contribute to this but 
also takes into account the freedom in specifying $\zeta$ on 
successive surfaces.  (Recall that $\zeta$ is determined only up to a 
fractional linear transformation by the geometry; it was suggested the 
remaining freedom in $\zeta$ be fixed by comparison with the numerical 
coordinate system in order to have as direct an interpretation as 
possible.  Cf. the paragraph following the one containing 
Eqn.~(\ref{ahol}).)

\subsubsection{The Conformal Structure of the Generators}

As emphasized above, because of finite-size effects, the conformal 
structure of the space of generators is {\em not} preserved under 
evolution.  The question of how severe this issue is governs the 
degree to which there is a sense of asymptotically constant spinors, 
vectors and tensors which is stable under evolution.  This 
subsubsection discusses the obstruction.

The space of generators has naturally the structure of a smooth 
oriented two-sphere, and a conformal structure on this is equivalent 
to a choice of complex structure, and this in turn is equivalent to 
giving a complex basis vector (or covector) up to proportionality.  In 
out case, we could take the basis vector to be $M^a$; then $\Lie _w 
M^a$, modulo terms proportional to $M^a$, would give a measure of the 
rate of 
change in complex structure.  Equivalently, the shear $M_a\Lie _wM^a$ 
of $M^a$ along $w^a$ would measure the rate of change.  This quantity 
can be readily computed but it is not directly useful here.

The more direct way of accounting for the change in complex structure 
is the part of $d\cz /d\eta$ (Eqn.~(\ref{dcza})) antiholomorphic in $
\zeta$; were $d\cz /d\eta$ holomorphic, the complex structure would be 
unchanged at first order in $\eta$ and $d\cz /d\eta$ would induce an 
infinitesimal Lorentz motion.

In practice, it is likely that $d\cz /d\eta =w^\zeta$ will 
be close to holomorphic.
We saw above that there are two contributions to it, one
$w^\zeta _{\rm h}=-(\eth w^{11'}+Xw^{11'}) \rho ^{-1}\delta
\overline\zeta$ depending on $\zeta$ intrinsic to $\ess (\eta )$ and 
one $w^\zeta _{\rm v}$ depending on the extension of $\zeta$ off $\ess 
(\eta )$.  For the intrinsic one, we have
$X\approx -(3\rho )^{-1}\Psi _1$, and so, if 
$\ess (\eta )$ is in fact within the peeling regime we will have $X
\sim O(R^{-3})$.  If the sphere is nearly round and $w^{11'}$ is 
nearly constant, then $w^\zeta _{\rm h}$ will be small.  

The quantity $w^\zeta _{\rm v}$ represents the
rate of change of the stereographic coordinate as one 
moves normal to $\ess (\eta )$ along $w^a$.  
This means that
$w^\zeta _{\rm v}$ 
depends on just how 
$\zeta$ is extended off $\ess (\eta )$, 
which in turn depends on 
how the 
underlying numerical coordinates extend to the future of $\ess (\eta )
$.  As typically the 
extraction surfaces are very distant, nearly round, and these features 
are reflected to good approximation in the numerical coordinates (and 
preserved under evolution), we 
expect that $w^\zeta _{\rm v}$ will give something which is 
close to an infinitesimal fractional linear transformation.

\subsubsection{The Strong Conformal Geometry}

The ``strong conformal geometry'' of null infinity links the scales of 
vectors along the generators with the scales of those transverse to 
the generators~\cite{PR1986}.  It can be characterized by the quantity
\begin{equation}\label{scg}
  \sqrt{(2i)^{-1}|P_{\rm B}|^{-2}d\zeta \wedge d{\overline\zeta}}
    \frac{\partial}{\partial u}\, ,
\end{equation}    
where the forms are defined on the space of tangent vectors to $\scrif 
(\eta )$ modulo the vectors tangent to the generators, and the square 
root is taken in the sense of a line-bundle-valued quantity.  (The 
root taken is irrelevant; there are also other equivalent ways of 
characterizing the structure.)  This quantity is independent of the 
Bondi frame.

We have so far constructed, for each $\scrif (\eta )$, a stereographic
coordinate $\zeta _\eta$ on the space of its generators, and we have
transition functions $\zeta _\eta\circ \zeta _{\eta _0}^{-1}$ relating
these for different values of $\eta$.  On any $\scrif (\eta )$ we may
define a Bondi coordinate $u_\eta$ with respect to the Bondi frame
defined by $\zeta _\eta$ (and $P_{\rm B}$), fixing the zero of $u_\eta 
$
to lie on the preferred cut.  Thus the quantity~(\ref{scg}) is
well-defined, and requiring it to be preserved under changes of $\eta$
leads to a transformation law for the vectors $\partial /\partial
u_\eta$.

We have
\begin{eqnarray}
 & (2i)^{-1} |P_{\rm B}|^{-2}d\zeta _\eta \wedge d\overline{\zeta _
\eta}  =(2i)^{-1}(1+|\zeta _\eta |^2)^{-1}d\zeta _\eta \wedge d
\overline{\zeta _\eta}\nonumber\\
 & =\frac{(1+|\zeta _{\eta _0}|^2)^2}{(1+|\zeta _{\eta}|^2)^2}
    \left|\frac{\partial\zeta _\eta}{\partial\zeta _{\eta _0}}\right|
^2
      (2i)^{-1}(1+|\zeta _{\eta _0} |^2)^{-1}d\zeta _{\eta _0} \wedge 
d\overline{\zeta _{\eta _0}}\, ,\qquad
\end{eqnarray}
and so we have 
\begin{equation}\label{utrans}
  \frac{\partial}{\partial u_\eta}
    =\frac{1+|\zeta _{\eta}|^2}{1+|\zeta _{\eta_0}|^2}
    \left|\frac{\partial \zeta _{\eta _0}}{\partial\zeta}\right|
  \frac{\partial}{\partial u_{\eta _0}}\, .
\end{equation}

\subsubsection{Identification of the Generators}

We now turn to the supertranslations identifying the generators of $
\scrif (\eta )$ for different values of $\eta$.  

We saw above (Eqn.~(\ref{uas}))
that $|\lambdabar |^2w^b l_b n^a_\eta$ represents the 
apparent displacement of the cut corresponding to $\ess (\eta +d\eta )
$ with respect to the preferred cut in $\scrif (\eta )$.  
To integrate these infinitesimal displacements, however, we must 
express them all as elements of the same vector space.  We may do this 
by using the transformation rule~(\ref{utrans}) 
relating $n^a_\eta =\partial /\partial u_\eta$ to $n^a_{\eta _0}=
\partial /\partial u_{\eta _0}$ derived above; the infinitesimal 
displacement, expressed as a vector at $\scrif (\eta _0)$ for some 
fixed $\eta _0$, is
$(1+|\zeta _{\eta _0}|^2)^{-1}(1+|\zeta _{\eta}|^2)|\partial \zeta 
_{\eta _0} /\partial \zeta _{\eta }||\lambdabar |^2w^bl_b  n^a_{\eta 
_0}$. 
Thus the supertranslation taking the cut labeled by $\eta _1$ to that 
labeled by $\eta _2$ will be, in $\scrif (\eta _0)$,
\begin{equation}
\Delta u _{\eta _0}(\eta _2,\eta _1)=\int _{\eta _1}^{\eta _2}
  \frac{1+|\zeta _{\eta}|^2}{1+|\zeta _{\eta _0}|^2}
  \left|\frac{\partial \zeta _{\eta _0}}{\partial \zeta _{\eta }}
\right| 
  |\lambdabar |^2w^al_a\, d\eta\, .
\end{equation}
In this integral, the coordinates $(\zeta _{\eta _0},\overline{\zeta 
_{\eta _0}})$ are held fixed and $\zeta _\eta =\cz (\zeta _{\eta _0},
\overline{\zeta _{\eta _0}})$.

The full transformation law for the Bondi retarded times will be, from 
Eqn.~(\ref{utrans}) again and this,
\begin{equation}
u_{\eta _2}=
\frac{1+|\zeta _{\eta _1}|^2}{1+|\zeta _{\eta_2}|^2}
    \left|\frac{\partial\zeta _{\eta _2}}{\partial\zeta _{\eta _1}}
\right|  u_{\eta _1} +\Delta u _{\eta _2}(\eta _2,\eta _1)\, .
\end{equation}    
One can verify directly that these transition functions are 
compatible, that is, computing $u_{\eta _3}$ either directly from 
$u_{\eta _1}$, or from $u_{\eta _2}$ in terms of $u_{\eta _1}$, gives 
the same answer.

What we shall actually need is to refer the angular momenta at 
different values of $\eta$ to a single value $\eta _0$.  We therefore 
define $\cu =u_\eta \circ u_{\eta _0}^{-1}\Bigr| _{\eta _0=0}=\Delta u 
_\eta 
(\eta ,\eta _0)$.  Explicitly,
\begin{equation}\label{cueq}
\cu (\eta _1)=\int _{\eta _0}^{\eta _1}
  \frac{1+|\zeta _{\eta}|^2}{1+|\zeta _{\eta _1}|^2}
  \left|\frac{\partial \zeta _{\eta _1}}{\partial \zeta _{\eta }}
\right| 
  |\lambdabar |^2w^al_a\, d\eta\, .
\end{equation}
In this integral, 
we have $\zeta _\eta =\cz (\zeta _0,\overline{\zeta _0},\eta _0,\eta )
$ in order to keep the generator of null infinity fixed.  This applies 
not only to the explicit factors of $\zeta _\eta$, $\zeta _{\eta _1}$, 
but also to the dependences $|\lambdabar |^2 w^al_a$ and the
derivative terms.
(So
$\partial \zeta _{\eta _1}/\partial \zeta _\eta =\left(\partial \zeta 
_\eta /\partial \zeta _{\eta _0}\right) ^{-1} \left(\partial \zeta 
_{\eta _1} /\partial \zeta _{\eta _0}\right) =\left(\partial\cz (\zeta 
_0,\overline{\zeta _0},\eta _0,\eta )/\partial \zeta _0\right) ^{-1}
\left(\partial\cz (\zeta _0,\overline{\zeta _0},\eta _1,\eta )/
\partial \zeta _0\right)$.)

\subsection{Evolution of Energy--Momentum}\label{CEM}

To compare the energy--momentum at $\ess (\eta )$ with that at $\ess 
(\eta _0)$, 
then, we express 
$P^{AA'}_{\ess (\eta )}{\overline\pi}_A\pi _{A'}$ by giving $\pi _{A'}
$ 
as a function of $\zeta$;  we then insert for this function $\cz$.  We 
have
\begin{widetext}
\begin{eqnarray} 
  P^{AA'}{\overline\pi}_A\pi _{A'}
  &=&P^{00'}{\overline\pi}_0\pi _{0'}
    +P^{01'}{\overline\pi}_0\pi _{1'}
    +P^{10'}{\overline\pi}_1\pi _{0'}
    +P^{11'}{\overline\pi}_1\pi _{1'}\nonumber\\
     &=&(1/2)(P^{00'}+P^{11'}) +(1/2)(P^{00'}-P^{11'}) \frac{1-|\cz |
^2}{1+|\cz |^2}
       -P^{01'}\frac{\overline\cz}{1+|\cz |^2}
       -P^{10'}\frac{\cz}{1+|\cz |^2}\, .\label{compenmom}
\end{eqnarray}
\end{widetext}
In this formula, the components of $P^a$ are evaluated at $\ess 
(\eta )$ 
as in Section~V.  The formula here gives the component of the 
energy--momentum at $\ess (\eta )$ in the direction specified by 
$\zeta$ at $\ess (\eta _0)$. 

As noted above, if the extraction surfaces $\ess (\eta )$ are far 
enough 
away that they provide good models of cuts of null infinity, then $\cz
$ 
will be a fractional linear transformation representing a Lorentz 
transformation and $P^{AA'}{\overline\pi}_A\pi _{A'}$ will be 
interpretable as the component of a covector along the null direction 
specified by $\zeta$.   Another way of saying this is that
energy--momentum~(\ref{compenmom}) will have only $j=0$ and $j=1$ 
components.   Because of finite-size effects, however, we cannot 
expect 
this to hold exactly, and there is a question of principle of how to 
extract the (co)vectorial part of the energy--momentum when these 
effects 
cannot be neglected.

The natural thing to do is find the boost relative to which the 
energy--momentum~(\ref{compenmom}) has zero dipole moment 
($j=1$ component), project the energy--momentum in this frame 
(that is, keep only the $j=0$ component in this frame), and then boost 
back to the asymptotic laboratory frame at $\ess (\eta _0)$.  (There 
will be a unique frame in which the dipole moment is 
zero~\cite{ADH1990}.)  While there is no simple closed-form expression 
for 
this, there is an iterative procedure which one would expect to 
converge rapidly.

The projection of the energy--momentum relative to the frame defined 
by a 
unit future-pointing vector $t^{AA'}$ will be 
\begin{equation}\label{itenmoma}
\begin{split}
P^{AA'}(t)=&2^{5/2}\pi ^{-1}\int\left[
    \begin{array}{cc} |\zeta |^2 &\zeta\\ \overline\zeta&1
  \end{array}\right](1-|\sigma |^2/\rho ^2) |\delta\zeta |^2\\
&\times  \frac{ P^{BB'}{\overline\pi}_B\pi _{B'}}{(t^{00'}-\overline
\zeta t^{01'}
     -\zeta t^{10'} +|\zeta |^2 t^{11'})^3} d\ess\, ,
\end{split}
\end{equation}
where $P^{BB'}{\overline\pi }_B\pi _{B'}$ is given by 
(\ref{compenmom}).  
Thus if, starting from any $t^{AA'}_0$, we define
\begin{equation}\label{itenmomb}
  t^{AA'}_{n+1}=(P^{00'}(t_n)P^{11'}(t_n)-|P^{01'}(t_n)|^2)^{-1} 
P^{AA'}(t_n)\, 
\end{equation}
then the sequence $t_n^{AA'}$ will converge to the time-direction 
determined 
by the Bondi--Sachs energy--momentum, with $P^{AA'}(t_n)$ converging 
to 
that energy--momentum.  If $\cz(\zeta ,\overline\zeta )$ is close to $
\zeta$ 
(as would often be expected), then it would be natural to choose 
$t_0^a =P^a/\sqrt{P_bP^b}$.  It may well not even be necessary to move 
to 
further $t_n^a$ in the sequence to attain the accuracy required in 
many situations.

\subsection{Evolution of Angular Momentum}\label{CAM}

The strategy for comparing the angular momenta at different cuts is 
similar 
to that for comparing the energy--momenta.  The essential difference 
is that 
we must refer all angular momenta to the same ``origin'' (that is, the 
same 
cross-section of $\scrif$).  We have already found that $\cu$ is the 
supertranslation relating the measurement of the angular momentum at 
$\ess (\eta )$ to that at $\ess (\eta _0)$.  Since a supertranslation 
acts on 
the twistors by simply being added to $\lambda _{\rm B}$ in the 
parameterization (\ref{Z0eq}), (\ref{Z1eq}), in order to refer the 
angular momentum back to the original cut $\scrC (\eta _0)$, we need 
to replace $\lambda _{\rm B}$ by $\lambda _{\rm B}-\cu$ in 
Eq.~(\ref{angmomeqzeta}), as well as replacing $\zeta$ by $\cz$.

Since, even at a single cut, the angular momentum in general 
relativity is 
given by an object with components for arbitrary $j\geq 1$, the 
angular 
momentum cannot be reduced to a vectorial object and so the sort of 
projection procedure which was used for the energy--momentum is not 
needed.  On the other hand, if it is desired to compute the components 
of 
$\mu$ for different values of $j$, this can certainly be done.  
However, 
there is no simple algebraic (that is, involving only finitely many 
operations for each term) transformation taking a resolution of $\mu$ 
in spherical harmonics at $\ess (\eta )$ (such as might be found via 
Eq.~(\ref{angmomsh}) and Tables~\ref{fifthtable}, \ref{sixthtable}) 
and transporting it to $\ess (\eta _0)$, even if $\cz (\zeta ,
\overline\zeta )$ 
is given by a fractional linear transformation, because the asymptotic 
laboratory frames at $\ess (\eta )$ and $\ess (\eta _0)$ might be 
relatively 
boosted, and boosts mix infinitely many $j$-values.  (Of course, if $
\cz$ 
can be approximated as differing from the identity only to a finite 
order, 
then an algebraic transformation can be derived.)

\subsection{Two Technical Points}\label{Tech}

The formulas derived above for the evolution of the energy--momentum 
and angular momentum were the present paper's goals.  As emphasized 
above, these results include possible finite-size corrections, which, 
if significant, should be interpreted as signs that the
extraction surfaces are not distant enough to give a stable model of
null infinity.  

I mention here two further issues related to these finite-size 
effects, points which do not figure in the results above but would be 
relevant if one were to try to draw broader lessons for the 
development of quasilocal kinematics from these results.

The first is that in general the comparisons of energy--momentum and 
angular momentum at $\ess (\eta _0)$ and $\ess (\eta _1)$ depend not 
just on these surfaces themselves but on the intermediate ones $\ess 
(\eta )$, $\eta _0\leq\eta\leq \eta _1$.  In other words, because of 
finite-size effects, one would not expect an {\em integrable} 
comparison.  This issue, of course, would disappear if the surfaces 
were actually at null infinity.

The second issue is that the structures discussed here do not actually 
determine how to evolve the phase of a twistor or spinor.  This does 
not lead to any difficulties in the formalism given here, but it would 
be a point to keep in mind in developing a more general theory.

\section{Users' Guide}\label{UG}

The preceding sections have covered the derivations of formulas for
the energy--momentum, angular momentum, and comparisons of them at
different times.  The aim of the present section is to give a users'
guide to the results.

The starting-point is a spacelike two-surface $\ess$ of spherical
topology (or, for evolution, a one-parameter family $\ess (\eta )$ of
such surfaces, with $\nabla _a\eta$ timelike).  We assume that a null
tetrad adapted to $\ess$ has been chosen, and the Newman--Penrose
quantities at $\ess$ are available in terms of this tetrad. 

The first step is to find a complex stereographic coordinate $\zeta$
on $\ess$; see the paragraph containing Eq.~(\ref{ahol}) and the two
paragraphs thereafter.  With this known, the factor $ \lambdabar$ 
giving
the rescaling to a Bondi--Sachs frame is determined by
Eq.~(\ref{lambdaeq}).  This defines an {\em asymptotic laboratory
frame}.

The second step is to compute the angular potential $\lambda _{\rm B}$
for the Bondi shear.  This may be done either via a Green's function
(using Eqs.~(\ref{Bshear}), (\ref{lgf}), (\ref{gf})), or by resolution
in spin-weighted spherical harmonics (eqs.~(\ref{shearresa}),
(\ref{shearresb}), (\ref{shearresc}); see also the last paragraph of
Section~\ref{BSF} for phase conventions).

The third step is to compute the components $A_{\alpha\beta}$ of the
kinematic twistor.  These are given by integrals over $\ess$.
Table~\ref{fourthtable} lists the integrands, which require, besides
the quantities already discussed, the asymptotic forms $\Psi _1^0$,
$\Psi _2^0$, $\Psi _3^0$ of the Weyl curvature components in the poor
man's no-incoming-radiation approximation; these asymptotic forms are
given in Table~\ref{firsttable} in terms of quantities on $\ess$.  (As
noted in the text, Table~\ref{firsttable} lists these under the
assumption $|\sigma |\ll {|\rho |}$ on $\ess$, which should 
be very good
for most purposes.  Section II shows how to compute them more 
accurately, 
if required.)

With the components of $A_{\alpha\beta}$ known, the Bondi--Sachs
energy--momentum may be read off directly in the asymptotic laboratory
frame:  see Eq.~(\ref{Benmom}).

The angular momentum is reported as a function $\mu (\zeta
,\overline\zeta )$ of the asymptotic direction with respect to
asymptotic reference frame, with $\mu+\overline\mu$ giving the
energy--moment in the direction and $i\mu-i\overline\mu$ the spatial
angular momentum about that axis.  One could choose to either present
this function directly or to give its resolution in spherical
harmonics.  For a direct presentation, the function is given by
Eq.~(\ref{angmomeqzeta}), and this can be resolved into spherical 
harmonics
by standard means.  If the components of $\lambda _{\rm B}$ in 
spherical 
harmonics have already been computed, then the resolution of $\mu$ is
given by Eq.~(\ref{angmomsh}), which makes use of Tables~ 
\ref{fifthtable}
and \ref{sixthtable}.

If the spin and center-of-mass are required, one must transform to a
boosted asymptotic frame in which the time axis lies along the
Bondi--Sachs energy--momentum~\cite{ADH2007}.  

To study the evolution of the energy--momentum and the angular
momentum, one must, besides computing them on the different surfaces
$\ess (\eta )$, also give an invariant method for relating the
quantities on one surface to those on another.  If $w^a$ is a vector
from $\ess (\eta )$ to $\ess (\eta +d\eta )$, one solves
Eqns.~(\ref{czeq}),
(\ref{cueq}),
to find the functions $\cz (\zeta ,\overline\zeta ,\eta )$ and
$\cu (\zeta ,
\overline\zeta ,\eta )$ 
expressing the appearance of the cut $\scrC (\ess (\eta ))$ relative 
to
$\scrC (\ess (\eta _0))$.  Then Eq.~(\ref{compenmom}) (with $P^{AA'}$
defined by (\ref{Benmom}) evaluated at $\ess (\eta )$) gives the
energy--momentum at $\ess (\eta )$ in the directions as specified by 
the
angular variables at $\ess (\eta _0)$.  This function may, because of
finite-size effects, in general not be simply a vector but will have
components in all $j\geq 0$ representations.  To project the
vectorial part invariantly, one solves~(\ref{itenmomb}),
(\ref{itenmoma}) iteratively.
The angular momentum $\mu$ at
$\ess (\eta )$ but referred to the Bondi coordinates constructed 
at $\ess (\eta _0)$ is given by replacing $\lambda _{\rm B}$ by 
$\lambda _{\rm B}-\cu$ and $\zeta$ by $\cz$ in 
Eq.~(\ref{angmomeqzeta}).  
There is no simple formula for the evolution of the components of the 
angular 
momentum in spherical harmonics in the most general case; one must 
compute the components from the evolved $\mu$ (cf. Section VII~D).

\appendix*
\section{Some Properties of Spin-Weighted Spherical Harmonics}

This paper relies on some technical properties of spin-weighted
spherical harmonics, which are derived here.  The conventions are
those of ref.~\cite{PR1984}.

\subsection*{Definitions}

The spin-weighted spherical harmonics are determined as follows.  Fix
a spin frame $\homi ^A$, $\hio ^A$ (normalized with $\homi _A\hio
^A=1$ and with $2^{1/2}t^{AA'}=\homi ^A\homi^{A'}+\hio ^A\hio ^{A'}$).  
Put 
\begin{equation}
  Z(j,m)_{B\ldots CD\ldots E}
  =\underbrace{\homi _{(B}\cdots\homi _C}_{j-m}\underbrace{\hio 
_D\cdots\hio _{E)}}_{j+m}\, .
\end{equation}
Now let $\omicron ^A$, $\iota ^A$ be a second normalized frame, which
is considered to vary and to determine a point on the sphere
(corresponding to the null vector $\omicron ^A\omicron ^{A'}$, say).
Then one puts
\begin{eqnarray}
  {}_sZ_{j,m}&=&
  Z(j,m)_{B\ldots E}\underbrace{\omicron ^B\cdots\omicron ^C}_{j+s}
\underbrace{\iota ^D\cdots\iota ^E}_{j-s}\, ,\\
  {}_sY_{j,m}&=&(-1)^{j+m}\ {}_sZ_{j,m}\nonumber\\
&\times & \sqrt{\frac{(2j+1)! (2j)!}{4\pi (j+s)! (j-s)! (j+m)! (j-
m)!}}
  \, .\qquad
\end{eqnarray}
One has $\overline{{}_sY_{j,m}}=(-1)^{m+s}{}_{-s}Y_{j,-m}$.

The spin-weighted spherical harmonics as defined above are functions
on certain line bundles over the sphere.  However, it is common to
represent them by ordinary functions, by giving their values on
preferred sections.  There are two main conventions for this.  In the
first, the spin-frame is adapted to the complex stereographic
coordinate $\zeta$ and given by
\begin{eqnarray}
  \omicron ^A (\zeta ,\overline\zeta )
  &=&i(1+|\zeta |^2)^{-1/2}(-\zeta\homi ^A -\hio ^A )\\
  \iota ^A (\zeta ,\overline\zeta )
  &=& i(1+|\zeta |^2)^{-1/2}(-\homi ^A +\overline\zeta \hio ^A)\, ;
\end{eqnarray}
in the second, the adaptation is to the polar coordinates $\theta$,
$\phi$, and
\begin{eqnarray}
\omicron ^A(\theta ,\phi )&=&e^{i\phi /2}\cos (\theta /2)\homi ^A
  +e^{-i\phi /2}\sin (\theta /2)\hio ^A\\
  \iota ^A(\theta ,\phi )
   &=&-e^{i\phi /2}\sin (\theta /2)\homi ^A+e^{-i\phi /2}\cos 
(\theta /2)\hio ^A  \, .
   \quad\qquad
\end{eqnarray}
These frames differ by a phase only; one has $\homi ^A(\theta ,\phi
)=-e^{-i\phi /2}\homi ^A(\zeta ,\overline\zeta ) =-(\overline\zeta /
\zeta
)^{1/2}\omicron ^A(\zeta ,\overline\zeta )$.  Note that this means 
that, viewed as
ordinary functions, we have
\begin{equation}
  {}_sY_{j,m}(\theta ,\phi )=(-(\overline\zeta /\zeta )^{1/2})^{2s}\,
{}_sY_{j,m}(\zeta ,\overline\zeta)\, .
\end{equation}

In this paper, the spin-weighted harmonics are applied at two stages.
First, they are used in the solution of the twistor equation; in this
case, the variable frame is determined by $\omicron ^{\rm B}_A$ (and
the fixed frame by the asymptotic reference frame).  The second
occurrence of the harmonics is in parameterizing the $\pi _{A'}$
spinors appearing in the definitions of energy--momentum and angular
momentum.  In those case, it is ${\overline\pi}_A$ which takes on the
role of the variable spinor $\omicron _A$ in the spherical harmonics.
We note that in this case
\begin{eqnarray}
  \pi _{0'}&=&\overline{{\overline\pi}_A\homi
^A}={}_{-1/2}Z_{1/2,-1/2}\\
  \pi _{1'}&=&\overline{{\overline\pi}_A\hio ^A}
  =-{}_{-1/2}Y_{1/2,1/2}
\end{eqnarray}
and similarly
\begin{eqnarray}
\pi _{0'}\pi _{0'}&=&{}_{-1}Z_{1,1}=\sqrt{4\pi /3}\,{}_{-1}Y_{1,1}   
\\
\pi _{0'}\pi _{1'}&=&-{}_{-1}Z_{1,0}=-\sqrt{4\pi /6}\, {}_{-1}Y_{1,0}
\\
\pi _{1'}\pi _{1'}&=&{}_{-1}Z_{1,-1}=\sqrt{4\pi /3}\,
{}_{-1}Y_{1,-1}\\
\pi _{0'}{\overline\pi}_1&=&{}_0Z_{1,1}=\sqrt{4\pi /6}\, {}_0Y_{1,1}\\
\pi _{0'}{\overline\pi}_{0'}-\pi _{1'}{\overline\pi}_1
  &=&2\, {}_0Z_{1,0}=\sqrt{4\pi /3}\, {}_0Y_{1,0}\\
\pi _{1'}{\overline\pi}_0&=&-{}_0Z_{1,-1}=-\sqrt{4\pi /6}\,
{}_0Y_{1,-1}
\, .\qquad
\end{eqnarray}

\subsection*{Behavior Under Inversion}

Each of the spin-weighted spherical harmonics is defined as a function
of the spinor $\omicron _A$.  We consider here how the harmonics
change when the spinor is acted on by a spatial inversion

There is a choice of sign in lifting the inversion
from vectors to
spinors; we use $\alpha _A\mapsto
2^{1/2}t_{AA'}{\overline\alpha}^{A'}$.  (Then the spinor
${\acute\pi}_{A'}$ appearing in the treatment of angular momentum is
the image of $\pi _{A'}$ under inversion.)

Each of
the harmonics is given as a (normalization factor times a) 
function $Z_{A\ldots CD\ldots F}\omicron
^A\cdots\omicron ^C\iota ^D\cdots\iota ^F$, 
where there are $j+s$ omicrons and $j-s$ iotas.  The
antipodal map 
gives $\omicron _A\mapsto -\iota
_A$, $\iota _A\mapsto\omicron _A$.  This will evidently effect a
change ${}_sY_{j,m}\mapsto (-1)^{j+s}{}_{-s}Y_{j,m}$.
(Reversing the sign in the definition of the antipodal map on spinors
would change the action on spin-weighted spherical
harmonics to ${}_sY_{j,m}\mapsto (-1)^{j-s}{}_{-s}Y_{j,m}$, that is,
would contribute an extra minus sign for half-integral spin weights.
In this paper, since all final quantities have integral spin-weights,
the sign convention, as long as it is kept fixed, is unimportant.)

However, when one represents the harmonics by ordinary functions,
their
behavior under inversions appears more complicated, because the 
sections
used to effect the trivialization of the bundles are not invariant
under inversions.  Indeed, evaluating the sections at the antipodal 
point (whose stereographic 
coordinate is $-{\overline\zeta}^{-1}$), we find
\begin{eqnarray}
\omicron ^A(-{\overline\zeta}^{-1},-\zeta ^{-1}) &=&-(\zeta /\overline
\zeta )^{1/2}\iota 
^A(\zeta ,\overline\zeta )\, ,\\
\iota ^A(-{\overline\zeta}^{-1},-\zeta ^{-1})&=& (\overline\zeta /
\zeta )^{1/2}\omicron 
^A(\zeta ,\overline\zeta )\, .
\end{eqnarray}
Thus 
\begin{equation}
{}_sY_{j,m}(-{\overline\zeta}^{-1} ,-\zeta ^{-1})
=(-1)^{j+s} (\zeta /\overline\zeta )^s\ {}_{-s}Y_{j,m}(\zeta ,
\overline\zeta )\, ,
\end{equation}
where the extra factor is the conversion from the gauge at one point 
to its antipodal point.

For the frame adapted to the polar system, one has
\begin{eqnarray}
\omicron ^A(\pi -\theta ,\phi +\pi )&=& -i\iota ^A(\theta ,\phi )\\
\iota  ^A(\pi -\theta ,\phi +\pi )&=&-i\omicron ^A(\theta ,\phi )\, ,
\end{eqnarray}
and so
\begin{equation}
{}_sY_{j,m}(\pi -\theta ,\phi +\pi ) =(-i)^{2j}\, {}_{-s}Y_{j,m}
(\theta ,\phi ) \, .
\end{equation}

\subsection*{Tensor Products}

We derive here the resolutions of certain products of spin-weighted
spherical harmonics (``Clebsch--Gordan decompositions'') which are
used in the text.  Because in all cases one of the factors has small
values of $j$ and $s$ (indeed, $j=0,1$, $s=-1,0,1$) it is easiest to
proceed iteratively.

Our starting-point is the identity
\begin{equation}
  \alpha _A\beta _{B\ldots DE}=\alpha _{(A}\beta _{B\ldots DE)}
    -\frac{2j}{2j+1}\epsilon _{A(B}\beta _{C\ldots E)F}\alpha ^F\, ,
\end{equation}
where $\alpha _A$ is any spinor and $\beta _{B\ldots DE}$ is any
totally symmetric spinor of valence $2j$.  Taking $\alpha _A=\homi _A$
or $\alpha =\hio _A$ and $\beta _{B\ldots E}=Z(j,m)_{C\ldots E}$, we
find
\begin{equation}
\begin{split}
Z(1/2,&-1/2)_AZ(j,m)_{B\ldots E}\\
  =&Z(j+1/2,m-1/2)_{AB\ldots E}  \\
  &+\frac{j+m}{2j+1}\epsilon _{A(B}Z(j-1/2,m-1/2)_{C\ldots E)}
\end{split}
\end{equation}  
and
\begin{equation}
\begin{split}
Z(1/2,&1/2)_AZ(j,m)_{B\ldots E}\\
=&Z(j+1/2,m+1/2)_{AB\ldots E}\\
& -\frac{j-m}{2j+1}\epsilon _{A(B}Z(j-1/2,m+1/2)_{C\ldots E)}\, .
\end{split}
\end{equation}
Contracting now with either $\omicron ^A$ or $\iota ^A$, and with
$\omicron ^B\cdots \omicron ^C$ ($j+s$ times) and $\iota ^D\cdots\iota
^E$ ($j-s$ times), we find
\begin{equation}
\begin{split}
{}_{1/2}&Z_{1/2,\pm 1/2} \, {}_sZ_{j,m}
  ={}_{s+1/2}Z_{j+1/2,m\pm 1/2} \\
  &\mp\frac{(j\mp m)(j-s)}{(2j)(2j+1)}\, {}_{s+1/2}Z_{j-1/2,m\pm 1/2}
\end{split}
\end{equation}
and
\begin{equation}
\begin{split}  
   {}_{-1/2}&Z_{1/2,\pm 1/2} \, {}_sZ_{j,m}
  ={}_{s-1/2}Z_{j+1/2,m\pm 1/2} \\
  &\pm\frac{(j\mp m)(j+s)}{(2j)(2j+1)}\, {}_{s-1/2}Z_{j-1/2,m\pm 1/2}
  \, .
  \end{split}
\end{equation}

We may by repeated application of these build up all the tensor
decompositions.  The cases we need are as follows.

We have ${}_{-1}Z_{1,0}={}_{-1/2}Z_{1/2,1/2}\, {}_{-1/2}Z_{1/2,-1/2}$.
Using this, we find, after some algebra
\begin{equation}
\begin{split}
{}_{-1}Z_{1,0}\, {}_1Z_{j,m}=&{}_0Z_{j+1,m} -\frac{m}{2j}\,
{}_0Z_{j,m}\\
  &-\frac{(j^2-m^2)(j+1)}{4j(4j^2-1)}\, {}_0Z_{j-1,m}\, .
\end{split}
\end{equation}
Similarly ${}_{-1}Z_{1,\pm 1}=({}_{-1/2}Z_{1/2,\pm 1/2})^2$, from
which
\begin{equation}
\begin{split}
{}_{-1}Z_{1,\pm 1}\, {}_1Z_{j,m}=&{}_0Z_{j+1,m\pm 1}
  \pm \frac{j\mp m}{2j}\, {}_0Z_{j,m\pm 1}\\
  &+\frac{(j+1)(j\mp m)(j\mp m-1)}{4j(4j^2-1)}\, {}_0Z_{j-1,m\pm 
1}\, .
\end{split}
\end{equation}
Using ${}_0Z_{1,\pm 1}={}_{1/2}Z_{1/2,\pm 1/2}\, {}_{-1/2}Z_{1/2,\pm
1/2}$ we find
\begin{equation}
{}_0Z_{1,\pm 1}\, {}_0Z_{j,m}={}_0Z_{j+1,m\pm 1}
  -\frac{(j\mp m)(j\mp m-1)}{4(4j^2-1)}\, {}_0Z_{j-1,m\pm 1}
\end{equation}
and, using ${}_0Z_{1,0}=(1/2)\, {}_{1/2}Z_{1/2,1/2}\,
{}_{-1/2}Z_{1/2,-1/2} +(1/2)\, {}_{-1/2}Z_{1/2,1/2}\,
{}_{-1/2}Z_{1/2,-1/2}$,
\begin{equation}
{}_0Z_{1,0}\, {}_0Z_{j,m}={}_0Z_{j+1,m}
  +\frac{j^2-m^2}{4j(4j^2-1)}\, {}_0Z_{j-1,m}\, .
\end{equation}

\bibliography{../references}

\end{document}